\documentclass[MSNbibl,seceqn,number,dvips]{arxstspdf}
\usepackage{mathrsfs}
\usepackage{graphicx}
\usepackage{flushend}
\usepackage{stfloats}

\volume{28}
\issue{3}
\pubyear{2013}
\firstpage{335}
\lastpage{359}
\doi{10.1214/13-STS422} 

\makeatletter
\newcommand{\eqref}[1]{(\ref{#1})}

\newproclaim{exe}{Example}[section]
\newtheorem{prp}{Proposition}[section]
\newproclaim{rmk}{Remark}[section]
\newcommand{\bfpi}{\bolds{\pi}}
\def\simind{\stackrel{\mathrm{ind}}{\sim}}
\def\simiid{\stackrel{\mathrm{i.i.d.}}{\sim}}
\newcommand{\Unif}{\operatorname{Uniform}}
\makeatother

\begin{document}
\begin{frontmatter}

\title{MCMC for Normalized Random Measure Mixture Models}
\runtitle{MCMC for normalized random measure mixture models}

\begin{aug}
\author[a]{\fnms{Stefano} \snm{Favaro}\corref{}\thanksref{t1}\ead[label=e1]{stefano.favaro@unito.it}}
\and
\author[b]{\fnms{Yee Whye} \snm{Teh}\ead[label=e2]{y.w.teh@stats.ox.ac.uk}}
\runauthor{S. Favaro and Y.~W. Teh}
\affiliation{University of Torino and University of Oxford}

\address[a]{Stefano Favaro is Assistant Professor of Statistics, Department of Economics and
Statistics, University of Torino, C.so Unione Sovietica 218/bis, 10134 Torino, Italy \printead{e1}.}
\address[b]{Yee Whye Teh is Professor of Statistical Machine Learning, Department of Statistics, University of Oxford, 1 South
Parks Road, Oxford OX13TG, United Kingdom \printead{e2}.}

\thankstext{t1}{Also affiliated with Collegio Carlo Alberto, Moncalieri, Italy.}
\end{aug}

%
\begin{abstract}
This paper concerns the use of Markov chain Monte Carlo methods for
posterior sampling in Bayesian nonparametric mixture models with
normalized random measure priors. Making use of some recent posterior
characterizations for the class of normalized random measures,
we propose novel Markov chain Monte Carlo methods of both marginal type
and conditional type. The proposed marginal samplers are generalizations
of Neal's well-regarded Algorithm 8 for Dirichlet process mixture
models, whereas the conditional sampler is a variation of those
recently introduced in the
literature. For both the marginal and conditional methods, we consider
as a running example a mixture model
with an underlying normalized generalized Gamma process prior, and
describe comparative simulation results demonstrating the efficacies of
the proposed methods.
\end{abstract}

%
\vspace*{-6pt}

\end{frontmatter}

\section{Introduction}\label{intro}

Mixture models provide a statistical framework for modeling data where
each observation is assumed to have arisen from one of $k$ groups, with
$k$ possibly unknown, and each group being suitably modeled by a
distribution function from some parametric family. The distribution
function of each group is referred to as a component of the mixture
model and is weighted by the relative frequency of the group in the
population. Specifically, assuming $k$ being fixed, a collection of
observations $(Y_{1},\ldots,Y_{n})$ is modeled as independent draws
from a mixture distribution function with $k$ components, that is,
%
%
\begin{equation}
\label{eq:mod1} Y_{i} \simind \sum_{j=1}^{k}
\tilde{J}_{j}f(\cdot| \tilde{X}_{j}),
\end{equation}
where $f(\cdot| \tilde{X})$ is a given parametric family of
distribution functions indexed by a parameter $\tilde{X}$ and $(\tilde
{J}_{1},\ldots,\tilde{J}_{k})$ are the mixture proportions constrained
to be nonnegative and sum to unity. A convenient formulation of the
mixture model \eqref{eq:mod1} can be stated in terms of latent
allocation random variables, namely, each observation $Y_{i}$ is
assumed to arise from a specific but unknown component $Z_{i}$ of the
mixture model. Accordingly, an augmented version of \eqref{eq:mod1} can
be written in terms of a collection of latent random variables
$(Z_{1},\ldots,Z_{n})$, independent and identically distributed with
probability mass function $\mathbb{P}[Z_{i}=j]=\tilde{J}_{j}$, such
that the observations are modeled as
%
%
\begin{equation}
\label{eq:mod2} Y_{i} | Z_{i} \simind f(\cdot|
\tilde{X}_{Z_{i}}).
\end{equation}
Integrating out the random variables $(Z_{1},\ldots,Z_{n})$ then yields
\eqref{eq:mod1}. In a Bayesian setting the formulation of the mixture
model \eqref{eq:mod2} is completed by specifying suitable prior
distributions for the unknown quantities that are objects of the
inferential analysis: the parameter $(\tilde{X}_{1},\ldots,\tilde
{X}_{k})$ and the vector of proportions $(\tilde{J}_{1},\ldots,\tilde
{J}_{k})$. We refer to the monographs by Titterington et al.~\cite
{titsmimak85} and McLachlan and Basford~\cite{mclbas88} for accounts on
mixture models with a fixed number of components.
Markov chain Monte Carlo (MCMC) methods for Bayesian analysis of
mixture models with a fixed number of components was presented in
Dielbot and Robert~\cite{DieRob94}.

As regards the general case where the number of components is unknown,
a direct approach has been considered in Richardson and Green~\cite
{RicGre1997a}, who modeled the unknown $k$ by mixing over the fixed $k$
case, and made a fully Bayesian inference using the reversible jump
MCMC methods proposed in Green~\cite{Gre1995a}. See also Stephens~\cite
{Ste00} and references therein for some developments on such an
approach, whereas different proposals can be found in the papers by
Mengersen and Roberts~\cite{MarRob96}, Raftery~\cite{Raf96} and Roeder
and Wasserman~\cite{RoeWas97}. An early and fruitful approach, still in
the context of mixture models with an unknown number $k$ of components,
was proposed in Escobar~\cite{Esc88} who treated the problem in a
Bayesian nonparametric setting by means of a prior distribution based
on the Dirichlet process (DP) of Ferguson~\cite{Fer1973a}. This
approach arises as a major development of some earlier results in Lo
\cite{Lo1984a} and it is nowadays the subject of a rich and active literature.

In this paper we deal with mixture models with an unknown number of
components. In particular, we focus on a Bayesian nonparametric
approach with the specification of a class of prior distributions
generalizing the DP prior. In the Bayesian nonparametric setting the
central role is played by a discrete random probability measure $\tilde
{\mu}$ defined on a suitable measurable space~$\mathbb{X}$, an example
being the DP, whose distribution acts as a nonparametric prior. The
basic idea is that since $\tilde\mu$ is discrete, it can be written as
\[
\tilde\mu=\sum_{j\geq1}\tilde J_{j}
\delta_{\tilde{X}_{j}},
\]
where $(\tilde{J}_{j})_{j\geq1}$ is a sequence of nonnegative random
weights that add up to one and $(\tilde{X}_{j})_{j\geq1}$ is a sequence
of $\mathbb{X}$-valued random locations independent of $(\tilde
{J}_{j})_{j\geq1}$. Given $\tilde\mu$ and a collection of continuous
observations $(Y_{1},\ldots,Y_{n})$, a Bayesian nonparametric mixture
model admits a hierarchical specification in terms of a collection of
independent and identically distributed latent random variables
$(X_1,\ldots,X_n)$.\break Formally,
%
%
\begin{eqnarray}
\label{eq:hierarchical_intro} Y_i | X_i &
\simind& F(\cdot| X_i),
\nonumber\\
X_i | \tilde{\mu} & \simiid& \tilde{\mu},
\\
\tilde{\mu}& \sim& P,
\nonumber
\end{eqnarray}
where $P$ denotes the nonparametric prior distribution and $F(\cdot|
X_{i})$ is a probability distribution parameterized by the random
variable $X_{i}$ and admitting a distribution function $f(\cdot|
X_{i})$. Note that, due to the discreteness of $\tilde{\mu}$, each
random variable $X_i$ will take on value $\tilde{X}_j$ with probability
$\tilde{J}_j$ for each $j\geq1$, and the hierarchical model \eqref
{eq:hierarchical_intro} is equivalent to saying that observations
$(Y_1,\ldots,Y_n)$ are independent and identically distributed
according to a probability distribution $F$ with random distribution function
%
%
\begin{equation}
\label{eq:hierarchical_intro_1} f(\cdot)=\int
_{\mathbb{X}} f(\cdot| x){\tilde\mu}(dx)=\sum
_{j\geq
1}\tilde{J}_{j}f(\cdot| \tilde{X}_{j}).
\end{equation}
This is a mixture of distribution functions with a countably infinite
number of components. The probability distribution $F(\cdot| X_{i})$
is termed the mixture kernel, whereas the underlying distribution $P$
is\break termed the mixing distribution or, alternatively, the mixing
measure. Note that, since $\tilde\mu$ is discrete, each pair of the
latent random variables $(X_1,\ldots,X_n)$ will take on the same value
with positive probability, with this value corresponding to a component
of the mixture model. In this way, the latent random variables allocate
the observations $(Y_{1},\ldots,Y_{n})$ to a random number of
components, thus naturally providing a model for the unknown number of
components. Under the assumption of $\tilde{\mu}$ being a Dirichlet
process, the model \eqref{eq:hierarchical_intro_1} was introduced by Lo
\cite{Lo1984a} and it is known in Bayesian nonparametrics as the DP
mixture model.

The reason of the success of the Bayesian nonparametric approach in the
analysis of mixture models, as pointed out in the paper by Green and
Richardson~\cite{Gre01}, is that it exploits the discreteness of
$\tilde
{\mu}$, thus providing a flexible model for clustering items of various
kinds in a hierarchical setting without explicitly specifying the
number of components. Bayesian nonparametrics is now the subject of a
rich and active literature spanning applied probability, computational
statistics and machine learning. Beyond mixture analysis, Bayesian
nonparametrics has been applied to survival analysis by Hjort~\cite
{Hjo1990a}, to feature allocation models by Griffiths and
Ghahramani~\cite{GriGha2011a} and Broderick et al.~\cite{BroJorPit2012a} and to
regression (see the monograph by Rasmussen and Williams~\cite
{RasWil2006a}), among others. The reader is referred to the
comprehensive monograph edited by Hjort et al.~\cite{HjoHolMul2010a}
for a collection of reviews on recent developments in Bayesian nonparametrics.

Several MCMC methods have been proposed for posterior sampling from the
DP mixture model. Early works exploited the tractable marginalization
of $\tilde\mu$ with respect to the DP mixing distribution, thus
removing the infinite-dimensional aspect of the inferential problem.
The main references in this research area are represented by the
sampling methods originally devised in Escobar~\cite{Esc88,Esc94},
MacEachern~\cite{Mac1994a} and Escobar and West~\cite{EscWes1995a}, and
by the subsequent variants proposed in MacEachern~\cite{Mac1998a} and
MacEachern and M\"uller~\cite{MacMul1998a}. In Bayesian nonparametrics
these MCMC methods are typically referred to as marginal samplers and,
as noted by Ishwaran and James~\cite{IshJam2001a}, apply to any mixture
model for which the system of predictive distributions induced by
$\tilde{\mu}$ is known explicitly. The reader is referred to Neal
\cite
{Nea2000a} for a detailed overview of marginal samplers for DP mixture
models and for some noteworthy developments in this direction, such as
the well-known Algorithm 8 which is now a gold standard against which
other methods are compared.

An alternative family of MCMC methods for posterior sampling from the
DP mixture model is typically referred to as conditional samplers and
relies on the simulation from the joint posterior distribution,
including sampling of the mixing distribution $\tilde{\mu}$. These
methods do not remove the infinite-dimensional aspect of the problem
and instead focus on finding appropriate ways for sampling a finite but
sufficient number of the atoms of $\tilde{\mu}$. Ishwaran and James
\cite{IshJam2001a} proposed the use of a deterministic truncation level
by fixing the number of atoms and then bounding the resulting
truncation error introduced; the same authors also showed how to extend
the proposed method to any mixing distribution $\tilde{\mu}$ in the
class of the so-called stick-breaking random probability measures.
Alternatively, Muliere and Tardel\-la~\cite{multar98} proposed the use of
a random truncation level that allows one to set in advance the
truncation error. The idea of a random truncation has been recently
developed by Papaspiliopoulos and Roberts~\cite{PapRob2008a} who
proposed a Metropolis--Hastings sampling scheme, while Walker~\cite
{Wal2007a} proposed the use of a slice sampling scheme. See also
Papaspiliopoulos~\cite{Pap2008a} and Kalli et al.~\cite
{KalGriWal2011a} for further noteworthy improvements and developments
of conditional samplers with random truncation levels.

It is apparent that one can replace the DP mixing distribution with the
distribution of any other discrete random probability measure.
Normalized random measures (NRMs) form a large class of such random
probability measures. This includes the DP as a special case, and was
first proposed as a class of prior models in Bayesian nonparametrics by
Regazzini et al.~\cite{RegLijPru2003a}.
See also James~\cite{james}.
Nieto-Barajas et al.~\cite
{NiePruWal2004a} later proposed using NRMs as the mixing distribution
in \eqref{eq:hierarchical_intro_1}, while Lijoi et al.~\cite
{LijMenPru2005b,LijMenPru2005a,LijMenPru2007a} investigated explicit
examples of NRMs such as the generalized DP, the normalized $\sigma
$-stable process, the normalized inverse Gaussian process (NIGP) and
the normalized generalized Gamma process (NGGP).\break Various structural
properties of the class of NRMs have been extensively investigated by
James~\cite{Lan2003a}, Nieto-Barajas et al.~\cite{NiePruWal2004a},
James et al.~\cite{JamLijPru2006a,JamLijPru2009a,JamLijPru2010a} and
Trippa and Favaro~\cite{TriFav2011a}. Recently James et al.~\cite
{JamLijPru2009a} described a slightly more general definition of NRMs
in terms of the normalization of the so-called completely random
measures (CRMs), a class of discrete random measures first introduced
by Kingman~\cite{Kin1967a}. We refer to Lijoi and Pr\"unster~\cite
{LijPru2010a} for a comprehensive and stimulating overview of
nonparametric prior models defined within the unifying framework of CRMs.

In this paper we study MCMC methods of both marginal and conditional
types for posterior sampling from the mixture model \eqref
{eq:hierarchical_intro_1} with a NRM mixing distribution. We refer to
such a model as a NRM mixture model. Historically, the first MCMC
methods for posterior sampling from NRM mixture models are of the same
type as those proposed by MacEachern~\cite{Mac1994a} and Escobar and
West~\cite{EscWes1995a} for DP mixture models: they rely on the system
of predictive distributions induced by the NRM mixing distribution. See
James et al.~\cite{JamLijPru2009a} for details. Typically these methods
can be difficult to implement and computationally expensive due to the
necessary numerical integrations. To overcome this drawback, we propose
novel MCMC methods of marginal type for NRM mixture models. Our methods
are generalizations of Neal's celebrated Algorithm 8~\cite{Nea2000a} to
NRM mixture models, and represent, to the best of our knowledge, the
first marginal type samplers for NRM\vadjust{\goodbreak} mixture models that can be
efficiently implemented and do not require numerical integrations. As
opposed to MCMC methods of marginal type, conditional samplers for NRM
mixture models have been well explored in the recent literature by
Nieto-Barajas and Pr\"unster~\cite{NiePru2009a}, Griffin and Walker~\cite{GriWal2011a}, Favaro and Walker~\cite{Fav12} and Barrios et al.~\cite{BarLijNie2012a}. Here we propose some improvements to the
existing conditional slice sampler recently introduced by Griffin and
Walker~\cite{GriWal2011a}.

For concreteness, throughout the present paper we consider as a running
example the NGGP mixture model, namely, a mixture model of the form
\eqref{eq:hierarchical_intro_1} with the specification of a NGGP mixing
distribution. The NGGP is a recently studied NRM generalizing the DP
and featuring appealing theoretical properties which turns out to be
very useful in the context of mixture modeling. We refer to Pitman~\cite{Pit2003a},
Lijoi et al.~\cite{LijMenPru2007a,LijPruWal2008a} for an account on these properties with a view toward
Bayesian nonparametrics. In particular, the NGGP mixture model has been
investigated in depth by Lijoi et al.~\cite{LijMenPru2007a} who
proposed a comprehensive and comparative study with the DP mixture
model emphasizing the advantages of such a generalization.

The paper is structured as follows. Section~\ref{sec:nrm} introduces NRMs and
defines the induced class of NRM mixture models. In Section~\ref{MCMC_section} we
present the proposed MCMC methods, of both marginal type and
conditional type, for posterior sampling from NRM mixture models.
Section~\ref{sec4} reports on simulation results comparing the proposed methods
on a NRM mixture model with an underlying NGGP mixing distribution. A
final discussion is presented in Section~\ref{Discussion_Section}.


\section{Normalized Random Measures}\label{sec:nrm}

We review the class of NRMs with particular emphasis on their posterior
characterization recently provided by James et al.~\cite
{JamLijPru2009a}. Such a characterization will be crucial in Section~\ref{MCMC_section} for devising MCMC methods for posterior sampling
from NRM mixture models.

\subsection{Completely Random Measures}

\begin{figure}[b]

\includegraphics{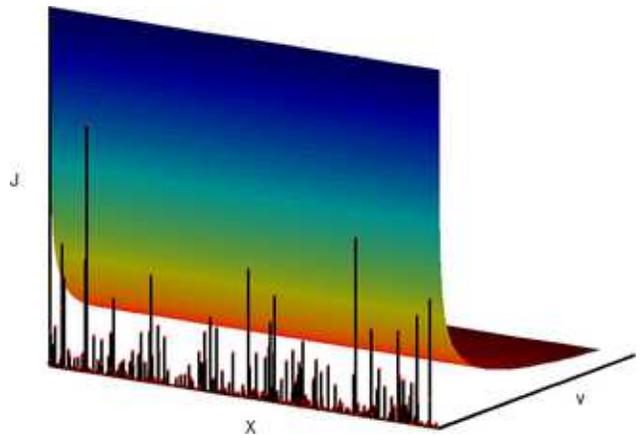}

\caption{A draw $\sum_{j\ge1} J_j \delta_{\tilde{X}_j}$ from a CRM.
Each stick denotes an atom in the CRM,
with mass given by its height $J_j$ and location given by $\tilde
{X}_j$. Behind the CRM is the density of its
L\'evy intensity measure $\nu$. The random point set $\{(J_j,\tilde
{X}_j)\}_{j\ge1}$ is described by a
Poisson process with intensity measure given by the L\'evy measure~$\nu
$.}\label{fig:crm}
\end{figure}

To be self-contained, we start with a description of CRMs. See the
monograph by Kingman~\cite{Kin1993a} and references therein for details
on such a topic.
Let $\mathbb{X}$ be a complete and separable metric space endowed with
the corresponding Borel $\sigma$-algebra $\mathscr{X}$.
A CRM on $\mathbb{X}$ is a random variable $\mu$ taking values on the
space of boundedly finite measures on $(\mathbb{X},\mathscr{X})$ and
such that for any collection of disjoint\vadjust{\goodbreak} sets $A_{1},\ldots,A_{n}$
in~$\mathscr{X}$, with $A_{i}\cap A_{j}=\varnothing$ for $i\neq j$, the
random variables ${\mu}(A_{1}),\ldots,{\mu}(A_{n})$ are mutually
independent. Kingman~\cite{Kin1967a} showed that a CRM can be
decomposed into the sum of three independent components: a nonrandom
measure, a countable collection of nonnegative random masses at
nonrandom locations and a countable collection of nonnegative random
masses at \mbox{random} locations. In this paper we consider CRMs consisting
solely of the third component, namely, a collection of random masses
$(J_{j})_{j\geq1}$ at random locations $(\tilde{X}_{j})_{j\geq1}$,
that is,
%
%
\begin{equation}
\label{eq:crm} \mu= \sum_{j\ge1} {J}_j
\delta_{\tilde{X}_j}.
\end{equation}
The distribution of $\mu$ can be characterized in terms of the
distribution of the random point set $(J_{j},\tilde{X}_{j})_{j\geq1}$
as a Poisson random measure on $\mathbb{R}^{+}\times\mathbb{X}$ with
mean measure $\nu$, which is typically referred to as the L\'evy
intensity measure. As an example, Figure~\ref{fig:crm} demonstrates a
draw of a CRM along with its L\'evy intensity measure.

%

For our purposes we focus on the so-called homogeneous CRMs, namely,
CRMs characterized by a L\'evy intensity measure $\nu$ factorizing as
$\nu(ds,dy)=\rho(ds)\mu_{0}(dy)$, for a nonnegative measure $\rho$
absolutely continuous with respect to Lebesgue measure and a nonatomic
probability measure $\mu_{0}$ over $(\mathbb{X},\mathscr{X})$. Such a
factorization implies the independence between the random masses
$(J_{j})_{j\geq1}$ and the random locations $(\tilde{X}_{j})_{j\geq1}$
in \eqref{eq:crm}. Hence, without loss of generality, the random
locations can be assumed to be independent and identically distributed\vadjust{\goodbreak}
according to the base distribution $\mu_{0}$, while the distribution of
the random masses $(J_{j})_{j\geq1}$ is governed by the L\'evy measure
$\rho$: it is distributed according to a Poisson random measure with
intensity $\rho$.

\subsection{Homogeneous Normalized Random Measures}

Homogeneous CRMs provide a fundamental tool for defining almost surely
discrete nonparametric priors via the so-called normalization approach.
Spe\-cifically, consider a homogeneous CRM $\mu$ with L\'evy intensity
measure $\nu(ds,dy)=\rho(ds)\mu_{0}(dy)$ and denote by $T=\mu
(\mathbb
{X})=\sum_{j\geq1}J_{j}$ the corresponding total mass. Then one can
define an almost surely discrete random probability measure on $\mathbb
{X}$ as follows:
%
%
\begin{equation}
\label{eq:series_normalized} \tilde{\mu}=\frac{\mu}{T}=\sum
_{j\geq1}\tilde{J}_{j}\delta _{\tilde{X}_{j}},
\end{equation}
where $(\tilde{J}_{j})_{j\geq1}$ is a sequence of random probabilities
defined by normalizing, with respect to $T$, the sequence of random
masses $(J_{j})_{j\geq1}$. To ensure that the
normalization in \eqref{eq:series_normalized} is a well-defined
operation, the random variable $T$ has to be positive and finite almost
surely; this is guaranteed by a well-known condition on the L\'evy
measure $\rho$, that is,
%
%
\begin{eqnarray}
\label{eq:condition} \int_{\mathbb{R}^{+}} \rho(ds)&=&+\infty,
\nonumber
\\[-8pt]
\\[-8pt]
\nonumber
 \int
_{\mathbb{R}^{+} } \bigl(1-e^{-s}\bigr) \rho(ds)&<&+\infty.
\end{eqnarray}
The random probability measure $\tilde{\mu}$ is known from James et al.
\cite{JamLijPru2009a} as a homogeneous NRM with L\'evy measure $\rho$
and base distribution $\mu_{0}$. See also Regazzini et al.~\cite
{RegLijPru2003a} for an early definition of NRMs. The idea of
normalizing CRMs, in order to define almost surely discrete
nonparametric priors, is clearly inspired by the seminal paper of
Ferguson~\cite{Fer1973a} who introduced the DP as a normalized Gamma CRM.
%
%
\begin{exe}[(DP)]
A Gamma CRM is a homogeneous CRM with L\'evy intensity measure of the form
\[
\rho_{a}(ds)\mu_{0}(dy)=as^{-1} e^{-s}\,ds
\mu_{0}(dy),
\]
where $a>0$. We denote a Gamma CRM by ${\mu}_{a}$ and its total
mass by $T_{a}$. Note that the L\'evy measure $\rho_{a}$ satisfies the
condition \eqref{eq:condition}, thus ensuring that the NRM
\[
\tilde{\mu}_{a}=\frac{{\mu}_{a}}{T_{a}}
\]
is a well-defined random probability measure. Specifically, $\tilde
{\mu
}_{a}$ is a DP with concentration parameter $a$ and base distribution
$\mu_{0}$.\vadjust{\goodbreak}
\end{exe}
Other examples of homogeneous NRMs have been introduced in the recent
literature. Notable among these in terms of both flexibility and
sufficient mathematical tractability is the normalized generalized
Gamma process (NGGP). Such a process, first introduced by Pitman~\cite
{Pit2003a} and then investigated in Bayesian nonparametrics by Lijoi et
al.~\cite{LijMenPru2007a}, is defined by normalizing the so-called
generalized Gam\-ma CRM proposed by Brix~\cite{Bri1999a}. Throughout this
paper we will consider the NGGP as a running example.
%
%
\begin{exe}[(NGGP)]
A generalized Gamma CRM is a homogeneous CRM with L\'evy intensity
measure of the form
%
%
\begin{eqnarray}
\label{eq:levy_gg} &&\rho_{a,\sigma,\tau}(ds)\mu_{0}(dy)
\nonumber
\\[-8pt]
\\[-8pt]
\nonumber
&&\quad=
\frac{a }{\Gamma(1-\sigma
)}s^{-\sigma-1}\mathrm{e}^{-\tau s}\,ds\,\mu_{0}(dy),
\end{eqnarray}
where $a>0$, $\sigma\in(0,1)$ and $\tau\geq0$. We denote a generalized
Gamma CRM by ${\mu}_{a,\sigma,\tau}$ and its total
mass by $T_{a,\sigma,\tau}$. Note that the L\'evy measure $\rho
_{a,\sigma,\tau}$ satisfies the condition \eqref{eq:condition}, thus
ensuring that the NRM
\[
\tilde{\mu}_{a,\sigma,\tau}=\frac{{\mu}_{a,\sigma,\tau
}}{T_{a,\sigma
,\tau}}
\]
is a well-defined random probability measure. Specifically, $\tilde
{\mu
}_{a,\sigma,\tau}$ is a NGGP with parameter $(a,\sigma,\tau)$ and base
distribution $\mu_{0}$.
\end{exe}

The NGGP includes as special cases most of the discrete random
probability measures currently applied in Bayesian nonparametric
mixture modeling. The DP represents a special case of a NGGP given by
$\tilde{\mu}_{a,0,1}$. Further noteworthy examples of NGGPs include:
the normalized $\sigma$-stable process, given by $\tilde{\mu
}_{a,\sigma
,0}$, first introduced by Kingman et al.~\cite{Kin1975a}
in relation to optimal storage problems, and the normalized inverse
Gaussian process (NIGP), given by $\tilde{\mu}_{a,1/2,\tau}$, recently
investigated by Lijoi et al.~\cite{LijMenPru2005a} in the context of
Bayesian nonparametric mixture modeling. As regards the celebrated
two-parameter Pois\-son--Dirichlet process, introduced by Perman\break
et al.~\cite{PerPitYor1992a}, this is not a NRM. However, it can be expressed
in terms of a suitable mixture of NGGPs. See Pitman and Yor~\cite
{PitYor1997a} for details on such a representation.

It is worth pointing out that the parameterization of the L\'evy
intensity measure \eqref{eq:levy_gg} is different from those proposed
in the past by Brix~\cite{Bri1999a}, Pitman~\cite{Pit2003a} and Lijoi
et al.~\cite{LijMenPru2007a}. Such a parameterization uses three
parameters rather than two parameters. This is so that our NGGP\vadjust{\goodbreak} can
easily encompass all the other NRMs mentioned above. The
three-parameter formulation does not lead to a strict generalization of
the two-parameter formulation since the $a$ and $\tau$ parameters are
in fact redundant. Indeed, rescaling $\mu_{a,\sigma,\tau}$ by a
constant $c>0$, which does not affect the resulting NRM, leads to a
generalized Gamma CRM with parameters $(ac^{\sigma},\sigma,\tau/c)$.

\subsection{Normalized Random Measure Mixture Models}\label{sec:nrmmixture}

Given a set of $n$ observations $\mathbf{Y}=(Y_{1},\ldots,Y_{n})$, a~NRM
mixture model consists of a corresponding set of latent random
variables $\mathbf{X}=(X_{1},\ldots,X_{n})$ conditionally independent
and identically distributed given a NRM mixing measure $\tilde{\mu}$.
According to the hierarchical formulation \eqref
{eq:hierarchical_intro}, a NRM mixture model can be stated as follows:
\begin{eqnarray}
Y_{i} | X_{i} & \simind& F(\cdot| X_i),\label{eq:nrmmixture}
\nonumber\\
X_i | \tilde{\mu} & \simiid&\tilde{\mu},
\nonumber
\\[-8pt]
\\[-8pt]
\nonumber
\tilde{\mu}& =& \frac{\mu}{T},
\nonumber
\\
\mu& \sim&\operatorname{CRM}(\rho,\mu_{0}),
\nonumber
\end{eqnarray}
where $\operatorname{CRM}(\rho,\mu_{0})$ denotes the law of the CRM
${\mu}$
with L\'evy measure $\rho$ and base distribution $\mu_0$. The rest of
this section elaborates on some posterior and marginal
characterizations for the NRM mixing measure $\tilde{\mu}$. These
characterizations will be useful in deriving the MCMC methods for
posterior sampling from the NRM mixture model \eqref{eq:nrmmixture}.

Because $\tilde{\mu}$ is almost surely discrete, ties may occur among
the latent random variables $\mathbf{X}$, so that $\mathbf{X}$ contains
$k\le n$ unique values. Hence, an equivalent representation of $\mathbf
{X}$ can be given in terms of the random partition on $[n]:=\{1,\ldots
,n\}$ induced by the ties and the unique values. Let $\bfpi$ be the
induced random partition of $[n]$, that is, a family of random subsets
of $[n]$ such that indices $i$ and $j$ belong to the same subset
(cluster) if and only if $X_i=X_j$. For each cluster $c\in\bfpi$, we
denote the corresponding unique value by $X^\ast_c$. In the context of
mixture modeling, the random partition $\bfpi$ describes the assignment
of observations to the various components, while the unique value
$X^\ast_c$ plays the role of the parameter associated with component $c$.

The random variables $\mathbf{X}$ are a sample from an exchangeable
sequence directed by $\tilde{\mu}$ and, accordingly, the induced random
partition $\bfpi$ is also exchangeable, namely, the probability mass
function of $\bfpi$ depends only on the number of clusters $|\bfpi|$
and the sizes of the clusters $\{|c|\dvtx c\in\bfpi\}$. Such a probability
mass function is known in the literature as the exchangeable partition
probability function (EPPF). See the monograph by Pitman~\cite
{Pit2006a} and references therein for details on this topic. The EPPF
induced by the NRM $\tilde{\mu}$ has been recently characterized by
James et al.~\cite{JamLijPru2009a} using an auxiliary random variable
$U$ whose conditional distribution, given the total mass $T$, coincides
with a Gamma distribution with shape $n$ and inverse scale $T$. In
particular, the joint conditional distribution of the random variables
$\mathbf{X}$ and $U$, given $\mu$, is
%
%
\begin{eqnarray}
\label{eq_jointpostU}&& \mathbb{P} \bigl[\bfpi=\pi,\bigl
\{X^\ast_c\in dx_c \dvtx c\in\pi \bigr\},U\in
du | \mu \bigr]
\nonumber
\\[-8pt]
\\[-8pt]
\nonumber
&&\quad= \frac{1}{\Gamma(n)}u^{n-1}e^{-Tu}\,du\prod
_{c\in
\pi} {\mu}(dx_c)^{|c|}.
\end{eqnarray}
The next propositions briefly summarize the posterior characterizations
introduced by James et al.~\cite{JamLijPru2009a}. We start by considering
the characterization of the EPPF and the system of predictive
distributions induced by a NRM $\tilde{\mu}$. Note that such a
characterization can
be derived from the distribution \eqref{eq_jointpostU} by means of an
application of the so-called Palm formula for CRMs. See, for example,
Daley and
Vere-Jones~\cite{DalVer2008a}.

%
\begin{prp}\label{thm_partition}
Let $\tilde{\mu}$ be a homogeneous NRM with L\'evy measure $\rho$ and
base distribution $\mu_{0}$. The induced joint distribution of
$\mathbf
{X}$ and $U$, with $\tilde\mu$ mar\-ginalized out, is given by
%
%
\begin{eqnarray}
\label{eq:joint_eppf} &&\mathbb{P} \bigl[\bfpi=\pi,\bigl
\{X^\ast_c\in dx_c \dvtx c\in\pi\bigr\},U\in
du \bigr]
\nonumber
\\[-8pt]
\\[-8pt]
\nonumber
&&\quad=\frac{1}{\Gamma(n)}u^{n-1}e^{-\psi(u)}\,du\prod
_{c\in\pi}\kappa _{|c|}(u)\mu_0(dx_c),
\end{eqnarray}
where $\psi(\cdot)$ denotes the Laplace exponent of the underlying CRM
$\mu$ and $\kappa_m(u)$ denotes the $m$th moment of the exponentially
tilted L\'evy measure\break $e^{-us}\rho(ds)$, that is,
%
%
\begin{eqnarray}
\label{laplace} \psi(u)&=&\int_{\mathbb{R}^{+} }\bigl(1-
\mathrm{e}^{-us}\bigr)\rho(ds),
\nonumber
\\[-8pt]
\\[-8pt]
\nonumber
 \kappa _{m}(u)&=&\int
_{\mathbb{R}^{+} } s^{m}{e}^{-us}\rho(ds).
\end{eqnarray}
In particular, by marginalizing out the auxiliary random variable $U$,
the EPPF of $\bfpi$ has the following expression:
\[
\mathbb{P} [\bfpi=\pi ] = \int_{\mathbb{R}^+} \frac
{1}{\Gamma
(n)}u^{n-1}e^{-\psi(u)}
\prod_{c\in\pi}\kappa_{|c|}(u) \,du,
\]
while the unique values $\{X_c^{*} \dvtx c\in\bfpi\}$ are independent and
identically distributed according to $\mu_0$. Together these
characterize the joint distribution of the latent variables $\mathbf{X}$.
Accordingly,
\begin{eqnarray*}
&&\mathbb{P}[X_{n+1}\in dx | U,\mathbf{X}]\\
&&\quad\propto\kappa_{1}(U)
\mu _{0}(dx)+\sum_{c\in\bfpi}\frac{\kappa_{|c|+1}(U)}{\kappa_{|c|}(U)}
\delta_{X^{\ast}_{c}}(dx)
\end{eqnarray*}
is the predictive distribution for a new sample\break $X_{n+1}\sim\tilde
{\mu
}$, given $U$ and $\mathbf{X}$ and once $\tilde\mu$ is marginalized out.
\end{prp}
Note that from the probability distribution \eqref{eq:joint_eppf}
follows the posterior distribution of $U$ given $\mathbf{X}$, that is,
%
%
\begin{equation}
\label{eq:cond_latent} \mathbb{P} [U\in du | \mathbf{X} ]\propto
u^{n-1}\mathrm {e}^{-\psi(u)}\,du\prod_{c\in\bfpi}
\kappa_{|c|}(u).
\end{equation}
The next proposition completes the posterior characterization for NRMs
by showing that the posterior distribution of a homogeneuos CRM $\mu$,
given $\mathbf{X}$ and~$U$, is still a CRM.

%
\begin{prp}\label{th_posterior}
Let $\tilde{\mu}$ be a homogeneous NRM with L\'evy measure $\rho$ and
base distribution $\mu_{0}$. The posterior distribution of the
underlying homogeneous CRM $\mu$, given $\mathbf{X}$ and $U$,
corresponds to
%
%
\begin{equation}
\label{eq:posterior} \mu| U,\mathbf{X}\sim{\mu^{\prime}}+\sum
_{c\in\bfpi}J^{\prime
}_{c}\delta_{X^{\ast}_{c}},
\end{equation}
where $\mu^{\prime}$ is a homogeneous CRM with an exponential tilted
L\'
evy intensity measure of the form
\[
\nu^{\prime}(ds,dy)=\mathrm{e}^{-Us}\rho(ds)\mu_{0}(dy)
\]
and where the random masses $\{J^{\prime}_{c}\dvtx c\in\bfpi\}$ are
independent of $\mu^{\prime}$ and among themselves, with conditional
distribution
\[
\mathbb{P} \bigl[J^{\prime}_c\in ds | U,\mathbf{X} \bigr]=
\frac
{1}{\kappa_{|c|}(U)}s^{|c|}\mathrm{e}^{-Us}\rho(ds).
\]
The posterior distribution of the NRM $\tilde{\mu}$, given $\mathbf{X}$
and~$U$, follows by normalizing the CRM $\mu| U,\mathbf{X}$.
\end{prp}

We conclude this section by illuminating Propositions \ref
{thm_partition} and~\ref{th_posterior} via their applications to the DP
and NGGP.

%
\begin{exe}[(DP)]
An application of Proposition~\ref{thm_partition} to the L\'evy measure
of the Gamma CRM shows that $\bfpi$ is independent of $U$, and its
distribution coincides with
%
%
\begin{eqnarray}
\label{eq:crp} \mathbb{P} [\bfpi=\pi| U ]&=&\mathbb{P} [\bfpi=\pi ]
\nonumber
\\[-8pt]
\\[-8pt]
\nonumber
&=&
\frac{\Gamma(a)a^{|\pi|}}{\Gamma(a+n)} \prod_{c\in\pi}\Gamma\bigl(|c|\bigr).
\end{eqnarray}
The corresponding predictive distributions are also independent of $U$
and are of the form
%
%
\begin{equation}
\label{eq:dppredictive} X_{n+1} | U,\mathbf{X}\sim\frac{a}{a+n}
\mu_{0}+\sum_{c\in\bfpi
}\frac
{|c|}{a+n}
\delta_{X_{c}^{\ast}}.
\end{equation}
An application of Proposition~\ref{th_posterior} shows that the
posterior distribution of $\mu$, given $U$ and $\mathbf{X}$,
corresponds to \eqref{eq:posterior} with $\mu'$ a Gamma CRM with L\'evy
intensity measure
\[
\nu^{\prime}(ds,dy)=a s^{-1}\mathrm{e}^{-s(U+1)}\,ds
\,\mu_{0}(dy),
\]
and random masses $J'_c$ distributed according to a Gamma distribution
with parameter $(|c|,U+1)$. Normalizing the posterior CRM, the
resulting posterior random probability measure $\tilde{\mu} |
U,\mathbf{X}$ does not depend on the scale $U+1$ and is still a DP,
with updated base measure
\[
\mu_{n}=a\mu_{0}+\sum_{c\in\bfpi}|c|
\delta_{X_{c}^{\ast}}.
\]
\end{exe}
The law of the random partition $\bfpi$ induced by the predictive
distributions \eqref{eq:dppredictive} is popularly known as the Chinese
restaurant process. The metaphor is that of a sequence of customers
entering a Chinese restaurant with an infinite number of round tables.
The first customer sits at the first table, and each subsequent
customer joins a new table with probability proportional to $a$, or a
table with $m$ previous customers with probability proportional to $m$.
After $n$ customers have entered the restaurant, the seating
arrangement of customers around tables corresponds to the partition
$\bfpi$, with probabilities given by \eqref{eq:crp}. Relating to
$\mathbf{X}$, each table $c\in\bfpi$ is served a dish $X^\ast_c$, with
$X_i=X^\ast_c$ if customer $i$ joined table $c$, that is, $i\in c$. See
Blackwell and MacQueen~\cite{BlaMac1973a} for a first characterization
of the predictive distributions \eqref{eq:dppredictive}. See also
Aldous~\cite{Ald1985a} for details and Ewens~\cite{Ewe1972a} for an
early account in population genetics.

%
\begin{figure*}

\includegraphics{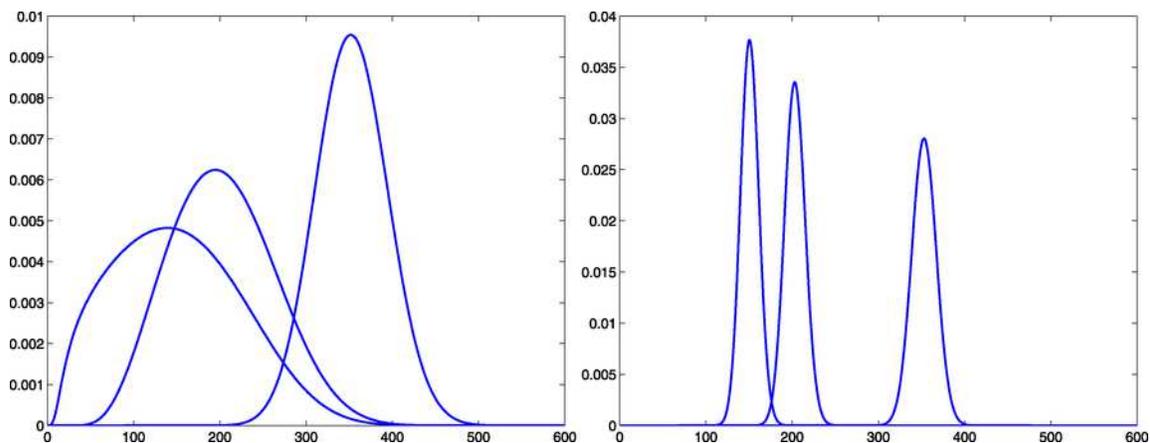}

\caption{Left: prior distribution of the number of clusters with
$\sigma
=0.7$, $\tau=1$, $a=0.1, 1$ and $10$ and $n=1000$. With increasing $a$
the number of clusters increases. Right: distribution of the number of
clusters with $\sigma=0.1$, $\tau=1$, and $a=38.5, 61.5$ and $161.8$.
Values of $a$ were chosen so that the mean number of clusters matches
those in the left panel. With a smaller value of $\sigma$ both the mean
and the variance in the number of clusters decreases, which is why the
values of $a$ are increased from the left panel.}\label{fig:distnumclusters}
\end{figure*}

%
\begin{exe}[(NGGP)]
An application of the formulae \eqref{laplace} to the L\'evy measure of
the generalized Gamma CRM leads to
%
%
\begin{eqnarray}
\label{eq:nggp_kappa} \psi(u)&=& \frac{a}{\sigma}\bigl((u+
\tau)^{\sigma}-\tau^\sigma\bigr),
\nonumber
\\[-8pt]
\\[-8pt]
\nonumber
 \kappa _m(u)&=&
\frac{a}{(u+\tau)^{m-\sigma}}\frac{\Gamma(m-\sigma
)}{\Gamma
(1-\sigma)}.
\end{eqnarray}
The random partition $\bfpi$ and $U$ are not independent as in the DP,
and has a joint distribution given by
%
%
\begin{eqnarray}
\label{eq:condeppf}\qquad &&\mathbb{P} [\bfpi=\pi,U\in du ]
\nonumber
\\
&&\quad = \frac{a^{|\pi
|}u^{n-1}}{\Gamma(n)(u+\tau)^{n-\sigma|\pi|}}
e^{-({a}/{\sigma})((u+\tau)^{\sigma}-\tau^\sigma)}\,du\\
&&\qquad{}\cdot
\prod_{c\in\pi}\frac{\Gamma(|c|-\sigma)}{\Gamma(1-\sigma)},\nonumber
\end{eqnarray}
and the corresponding system of predictive distributions for $X_{n+1}$,
given $U$ and $\mathbf{X}$, is
%
%
\begin{eqnarray}
\label{cond_pred} &&X_{n+1} | U,\mathbf{X}\nonumber\\
&&\quad\sim
\frac{a(U+\tau)^\sigma}{a(U+\tau)^\sigma+n-\sigma|\bfpi|} \mu _{0}
\\
&&\qquad{}+\sum_{c\in\bfpi}
\frac{|c|-\sigma}{a(U+\tau)^\sigma+n-\sigma|\bfpi
|}\delta _{X_{c}^{\ast}}.\nonumber
\end{eqnarray}
Finally, an application of Proposition~\ref{th_posterior} shows that
the posterior distribution of $\mu$, given $U$ and $\mathbf{X}$,
corresponds to
%
%
\begin{equation}
\label{eq:post_gg} \mu| U,\mathbf{X}\sim{\mu^{\prime}}+\sum
_{c\in\bfpi}J^{\prime
}_{c}
\delta_{X^{\ast}_{c}},
\end{equation}
where $\mu^{\prime}$ is a generalized Gamma CRM with parameters
$(a,\sigma,U+\tau)$ and the random masses $J_{c}^{\prime}$ are
independent among themselves and
of $\mu^{\prime}$, and distributed according to a Gamma distribution
with parameter $(|c|-\sigma,U+\tau)$.
\end{exe}

%
\begin{figure*}

\includegraphics{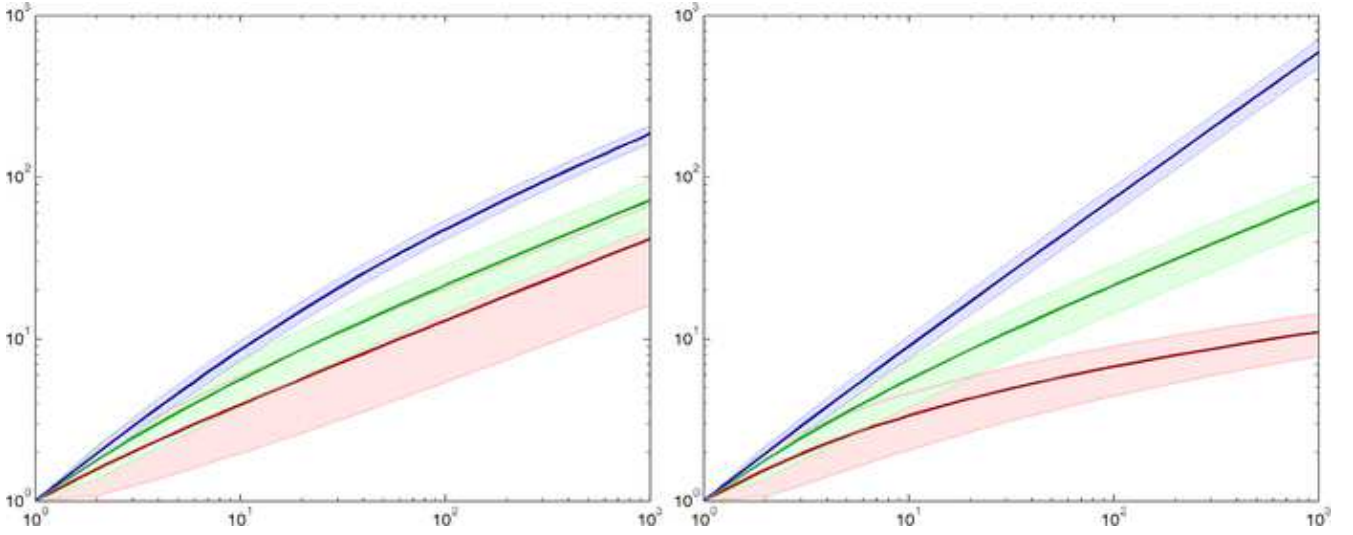}

\caption{Mean and standard deviation of the number of clusters as a
function of $n$, on a log--log plot. Left: with parameters $\sigma=0.5$,
$\tau=1$ and $a=0.1,1$ and $10$. Right: with parameters $\sigma=0.1,0.5$
and $0.9$, $\tau=1$ and $a=1$. The growth rate with $n$ follows a
power-law with index~$\sigma$, while $a$ affects the number of clusters
without affecting the power-law behavior.}\label{fig:growthnumclusters}
\end{figure*}

Note that the predictive distributions \eqref{cond_pred} provide a
generalization of the Chinese\vadjust{\goodbreak} restaurant process metaphor for the DP.
Conditionally on $U$, the probability of the $(n+1)$st customer joining
a table with $m$ existing customers is proportional to $m-\sigma$, with
$\sigma$ acting as a discount parameter. Note that the relative effect
of $\sigma$ is more pronounced for small values of $m$, which leads to
larger proportions of small tables with larger $\sigma$ and power-law
behaviors in $\bfpi$. On the other hand, the probability of joining a
new table is proportional to an increasing function of all three
parameters. Figure~\ref{fig:distnumclusters} shows how the distribution
over the number of clusters is affected by the parameters, while Figure~\ref{fig:growthnumclusters} shows how the distribution over the number
of clusters grows with $n$ for different values of the
parameters.

Lijoi et al.~\cite{LijMenPru2007a} provided a detailed comparative
study between the predictive structures of the NGGP and the DP in the
context of mixture modeling. The advantage of specifying the NGGP
mixing distribution with respect to the DP mixing distribution clearly
relies on the availability of the additional parameter $\sigma$. In the
DP mixture model the only free parameter which can be used to tune the
distribution of the number of clusters is the mass parameter $a$: the
bigger $a$, the larger the expected number of clusters. In the NGGP
mixture model the parameters $a$ and $\tau$ play the same role as the
mass parameter $a$ in the DP mixture model. On the other hand, $\sigma$
influences the grouping of the observations into distinct clusters and
can be used to tune the variance of the number of clusters in the NGGP
mixture model: the bigger $\sigma$, the larger the variance of the
number of clusters. Further, $\sigma$ also controls an interesting
reinforcement\vadjust{\goodbreak} mechanism that tends to reinforce significantly those
clusters having higher frequencies. This turns out to be a very
appealing feature in the context of mixture modeling. We refer to Lijoi
et al.~\cite{LijMenPru2007a} for details on the prior elicitation for
$\sigma$ to control the reinforcement mechanisms induced by it.


\section{MCMC Posterior Sampling Methods}\label{MCMC_section}

In this section we develop some novel MCMC samplers of both marginal
and conditional type for the NRM mixture models \eqref{eq:nrmmixture}.
In particular, we consider as a running example the NGGP mixing measure
with parameter $(a,\sigma,\tau)$ and base distribution~$\mu_{0}$.

\subsection{Conjugate Marginalized Sampler} \label{sec:conjmarg}

We start with the simplest situation, when the base distribution $\mu
_0$ is conjugate to the mixture kernel $F$. In this case both the CRM
$\mu$ and the cluster parameters $\{X^\ast_c\dvtx  c\in\bfpi\}$
can be marginalized out efficiently, leaving only the partition $\bfpi$
and auxiliary variable $U$ to be sampled. The joint distribution of
$\bfpi$ and $U$ is given by~\eqref{eq:joint_eppf}, while the
likelihood is
%
%
\begin{equation}
\label{eq:lik} \mathbb{P} [\mathbf{Y} | \bfpi=\pi ] = \prod
_{c\in\pi} f(\mathbf{Y}_c),
\end{equation}
where $\mathbf{Y}_c=\{Y_i\dvtx i\in c\}$ and
\[
f(\mathbf{Y}_c)= \int_{\mathbb{X} } \prod
_{i\in c} f(Y_i | x) \mu_{0}(dx).
\]
Since $\mu_0$ is conjugate to $F$, the integral is assumed to be
available in closed form and efficiently evaluated using the sufficient
statistics of $\mathbf{Y}_c$. Moreover, since both the conditional
distribution of $\bfpi$ given $U$ and the likelihood are in product
partition form, the conditional distribution of $\bfpi$ given $\mathbf
{Y}$ and $U$ is also in a product partition form.

We can update $\bfpi$ using a form of Gibbs sampling whereby the
cluster assignment of one data item $Y_i$ is updated at a time. Let
$\bfpi_{\setminus i}$ be the partition with $i$ removed.
We denote the cluster assignment of $Y_i$ with a variable $z_i$ such
that $z_i=c$ denotes the event that $Y_i$ is assigned to cluster $c\in
\bfpi_{\setminus i}$, and $z_i=\varnothing$ denotes the event that it is
assigned a new cluster. In order to update $z_i$, we can use formulae
\eqref{eq:joint_eppf} and \eqref{eq:lik} to provide the conditional
distribution of $z_i$, given $\bfpi_{\setminus i}$, $\mathbf{Y}$ and
$U$. Specifically,
%
%
\begin{eqnarray}
\label{eq:post_zi_1}\qquad &&\mathbb{P} [z_i=c |
\bfpi_{\setminus i},U,\mathbf{Y} ]
\nonumber
\\[-6pt]
\\[-6pt]
\nonumber
&&\quad\propto %
\cases{
\displaystyle\frac{\kappa_{|c|+1}(U)}{\kappa_{|c|}(U)} \frac{f(\{Y_i\}\cup\mathbf{Y}_c)}{f(\mathbf{Y}_c)}, &$\mbox{for $c\in \bfpi_{\setminus i}$,}$
\vspace*{3pt}
\cr
\kappa_1(U) f\bigl(\{Y_i\}\bigr), &$
\mbox{for $c=\varnothing$}$.} %
\end{eqnarray}
Under the assumption that $\tilde{\mu}$ is a NGGP and using~\eqref
{eq:nggp_kappa}, the above simplifies to
\begin{eqnarray*}
&&\mathbb{P} [z_i=c | \bfpi_{\setminus i},U,\mathbf{Y} ]
\\[2pt]
&&\quad\propto
\cases{\bigl (|c|-\sigma\bigr)f(Y_i | \mathbf{Y}_c),
&$\mbox{for $c\in\bfpi _{\setminus
i}$,}$ \vspace*{3pt}
\cr
a(U+
\tau)^{\sigma}f(Y_i | \varnothing), &$\mbox{for $c=
\varnothing$},$}
\end{eqnarray*}
where
\[
f(y | \mathbf{y})=\frac{f(\{y\}\cup\mathbf{y})}{f(\mathbf{y})}.
\]
We see that the update is a direct generalization of that for the DP
which can be easily recovered by setting $\sigma=0$. The probability of
$Y_i$ being assigned to a cluster is simply proportional to the product
of a conditional prior probability of being assigned to the cluster and
a conditional likelihood associated with the observation~$Y_i$. See
MacEachern~\cite{Mac1994a} and Neal~\cite{Nea1992b} for details on the
DP case. In the next section we describe the updates for the parameters
$a$, $\sigma$ and $\tau$, and for $U$, before proceeding to the
marginalized and conditional samplers in the case when $\mu_0$ is not
conjugate.

\subsubsection{Updates for NGGP parameters and $U$}\label{sec:hyperparam}

For~$U$, note that given $\bfpi$, $U$ is independent of $\mathbf{Y}$
with conditional distribution \eqref{eq:cond_latent}. In particular, in
the case of the NGGP, the
conditional distribution simplifies to
\[
\mathbb{P} [U\in du | \bfpi ] \propto\frac{u^{n-1}}{(u+\tau)^{n-a|\bfpi|}}e^{-({a}/{\sigma
})((u+\tau)^\sigma-\tau^\sigma)}\,du.
\]
A variety of updates can be used here. We have found that a change of
variable $V=\log(U)$ leads to better behaved algorithms, since the
conditional density $f_{V | \bfpi}(v)$ of $V$ given $\bfpi$, that is,
\begin{eqnarray*}
\mathbb{P} [V\in dv | \bfpi ]& \propto&\frac{e^{vn}}{(e^v+\tau)^{n-a|\bfpi|}}e^{-({a}/{\sigma
})((e^v+\tau)^\sigma-\tau^\sigma)}\,dv\\
& =&
f_{V|\bfpi}(v)\,dv,
\end{eqnarray*}
is log concave. We use a simple Metropolis--Hastings update with a
Gaussian proposal kernel with mean $V$ and variance $1/4$, although
slice sampling by\break Neal~\cite{Nea2003a} or, alternatively, adaptive
rejection sampling by Gilks and
Wild~\cite{GilWil1992a} can also be employed.

For the NGGP, we can easily derive the updates for the parameters $a$,
$\sigma$ and $\tau$ using \eqref{eq:condeppf} and given prior
specifications for the parameters. See Lijoi\break et~al.~\cite
{LijMenPru2007a} for a detailed analysis on prior specification in the
context of Bayesian nonparametric mixture modeling. As regards $a$, we
can simply use a Gamma prior distribution with parameter $(\alpha
_{a},\beta_{a})$. Then the conditional distribution of $a$, given
$\sigma$, $\tau$, $U$ and $\bfpi$, is simply a Gamma distribution,
that is,
\begin{eqnarray*}
&&\mathbb{P} [da | \sigma,\tau,U,\bfpi ]\\
&&\quad\propto a^{\alpha
_a+|\bfpi|-1}e^{-a (\beta_a+{((U+\tau)^\sigma-\tau
^\sigma
)}/{\sigma} )}\,da.
\end{eqnarray*}
For $\tau$ we can again use a Gamma prior distribution with parameter
$(\alpha_{\tau},\beta_{\tau})$.\vadjust{\goodbreak} Then the conditional distribution of
$\tau$, given
$a$, $\sigma$, $U$ and $\bfpi$, is
\begin{eqnarray*}
&&\mathbb{P} [d\tau| a,\sigma,U,\bfpi ] \\
&&\quad\propto \tau^{\alpha_\tau-1}e^{-\tau\beta_\tau}
\frac{e^{-
({a}/{\sigma})
((U+\tau)^\sigma-\tau^\sigma )}}{\tau^{\sigma|\bfpi
|}(U+\tau
)^{n-\sigma|\bfpi|}}\,d\tau.
\end{eqnarray*}
We update $\tau$ in its logarithmic domain, using the same procedure as
for $U$ described above. Finally, for $\sigma$ we can
use a Beta prior distribution with parameter $(\alpha_{\sigma},\beta
_{\sigma})$. Then the conditional distribution of $\sigma$, given $a$,
$\tau$, $U$ and $\bfpi$, corresponds to
\begin{eqnarray*}
&&\mathbb{P} [d\sigma| a,\tau,U,\bfpi ]\\
&&\quad \propto \sigma^{\alpha_\sigma-1}(1-
\sigma)^{\beta_\sigma-1} \frac{e^{-({a}/{\sigma})((U+\tau)^\sigma-\tau^\sigma)}}{\tau
^{\sigma
|\bfpi|}(U+\tau)^{n-\sigma|\bfpi|}}\\
&&\qquad{}\cdot\prod_{c\in\pi}
\frac{\Gamma
(|c|-\sigma)}{\Gamma(1-\sigma)}\,d\sigma.
\end{eqnarray*}
We can easily update $\sigma$ using slice sampling by\break Neal~\cite{Nea2003a}.

\subsection{Nonconjugate Marginalized Samplers}\label{sec:nonconjmarg}

The main drawback of the previous algorithm is the assumption of
conjugacy, which limits its applicability since nonconjugate priors are
often desirable in order to increase modeling flexibility. For DP
mixture models a number of marginalized algorithms for the nonconjugate
setting have been proposed and investigated in the literature. The
review of Neal~\cite{Nea2000a} provides a detailed overview along with
two novel algorithms. One of these algorithms, the so-called Algorithm
8, is simple to implement, has been demonstrated to provide excellent
mixing speed, and has a tunable parameter to trade off computation cost
against speed of convergence.

\subsubsection{Generalizing Neal's Algorithm 8}\label{sec:algorithm8}

In this section we provide a straightforward generalization of Neal's
Algorithm 8 to the class of NRM mixture models with a nonconjugate base
distribution. Here, the cluster parameters $X_c^\ast$ cannot be easily
marginalized out. Instead we include them into the state of the MCMC
algorithm, so that the state now consists of the partition $\bfpi$, $\{
X_c^\ast\dvtx  c\in\bfpi\}$ and the random variable~$U$, and we sample
the cluster parameters along with $\bfpi$ and~$U$. Note that the
parameters for existing clusters $X_c^\ast$ can be updated with
relative ease, using any MCMC update whose stationary distribution is
the conditional distribution of $X_c^\ast$ given everything else, that is,
\[
\mathbb{P} \bigl[X_c^\ast\in dx | \bfpi,U,\mathbf{Y}
\bigr] \propto\mu _0(dx) \prod_{i\in c}
f(Y_i | x) .
\]
The difficulty with a nonconjugate marginalized sampler is the
introduction of new clusters (along with their parameters) when Gibbs
sampling the cluster assignments. Following Neal~\cite{Nea2000a}, we
conceptualize our update in terms of an augmented state with additional
temporarily existing variables, such that the marginal distribution of
the permanent variables once the temporary ones are integrated out is
the appropriate posterior distribution.

Consider updating the cluster assignment variable $z_{i}$ given the
existing clusters in $\bfpi_{\setminus i}$. We introduce an augmented
space with $C$ empty clusters, with parameters $X_1^\mathrm{e},\ldots
,X_C^\mathrm{e}$ that are independent of $\bfpi_{\setminus i}$ and
independent and identically distributed according to $\mu_0$. The state
space of $z_i$ is augmented as well to include both existing clusters
$\bfpi_{\setminus i}$ and the new ones $[C]=\{1,\ldots,C\}$, with
conditional distribution
\[
\mathbb{P} [z_i=c\in\bfpi_{\setminus i} | \bfpi_{\setminus
i} ]
\propto\frac{\kappa_{|c|+1}(U)}{\kappa_{|c|}(U)}
\]
and
\[
\mathbb{P} \bigl[z_i=k\in[C] | \bfpi_{\setminus i} \bigr]\propto
\frac
{\kappa_{1}(U)}{C},
\]
respectively. Identifying $z_i$ being in any of the additional clusters
as assigning $Y_i$ to a new cluster, we see that the total probability
for $Y_i$ being assigned to a new cluster is proportional to the first
moment $\kappa_1(U)$, which is the same as in \eqref{eq:joint_eppf} and
\eqref{eq:post_zi_1}.

The update can be derived by first initializing the augmentation
variables given the current state of the Markov chain, updating $z_i$,
then discarding the augmentation variables. If $Y_i$ is currently
assigned to a cluster which contained another data item, then $z_i=c$
for some $c\in\bfpi_{\setminus i}$, and the empty cluster parameters
are simply drawn independently and identically according to $\mu_0$. On
the other hand, if $Y_i$ is currently assigned to a cluster containing
only itself, say, with parameter $X^\ast_\varnothing$, then in the
augmented space $z_i$ has to be one of the new clusters, say, $z_i=k$
for some $k\in[C]$ with $X^\mathrm{e}_k=X^\ast_\varnothing$. The actual
value of $k$ is unimportant, for convenience we may use $k=1$. The
other empty clusters then have parameters drawn independently and
identically according to $\mu_0$. We can now update $z_i$ by sampling
from its conditional distribution given $Y_i$ and the parameters of all
existing and empty clusters. Specifically,
%
%
\begin{eqnarray}
\label{eq:post_zi_2} &&\mathbb{P} [z_i=c |
\bfpi_{\setminus i},U,\mathbf{Y} ]
\nonumber
\\[-8pt]
\\[-8pt]
\nonumber
&&\quad\propto %
\cases{
\displaystyle\frac{\kappa_{|c|+1}(U)}{\kappa_{|c|}(U)} f\bigl(Y_i |
X_c^\ast\bigr),
&$\mbox{for $c\in\bfpi_{\setminus i}$,}$ \vspace *{2pt}
\cr
\displaystyle\frac{\kappa_{1}(U)}{C} f
\bigl(Y_i | X_c^\mathrm{e}\bigr), &$\mbox {for $c
\in[C]$}.$} %
\end{eqnarray}
Under the assumption that $\tilde{\mu}$ is a NGGP, \eqref{eq:post_zi_2}
again simplifies to
\begin{eqnarray*}
&&\mathbb{P} [z_i=c | \bfpi_{\setminus i},U,\mathbf{Y} ] \\
&&\quad\propto
\cases{ \bigl(|c|-\sigma\bigr) f\bigl(Y_i | X_c^\ast
\bigr), &$\mbox{for $c\in\bfpi_{\setminus
i}$,}$ \vspace*{2pt}
\cr
\displaystyle\frac{a}{C}(U+\tau)^\sigma f\bigl(Y_i |
X_c^\mathrm{e}\bigr), &$\mbox{for $c\in [C]$}.$} %
\end{eqnarray*}
If the new value of $z_i$ is $c\in[C]$, this means that $Y_i$ is
assigned to a new cluster with parameter $X_c^\mathrm{e}$; the other
empty clusters are discarded to complete the update. On the other hand,
if the new value is $c\in\bfpi_{\setminus i}$, then $Y_i$ is assigned
to an existing cluster~$c$, and all empty clusters are discarded.
Finally, the random variable $U$ and any hyperparameters may be updated
using those in Section~\ref{sec:hyperparam}.

\subsubsection{The Reuse algorithm}\label{sec:reuse}

In the above algorithm, each update to the cluster assignment of an
observation is associated with a set of temporarily existing variables
which has to be generated prior to the update and discarded afterward.
As a result, many independent and identically distributed samples from
the base distribution have to be generated throughout the MCMC run, and
in our experiments this actually contributes a significant portion of
the overall
computational cost.
We can mitigate this wasteful generation and discarding of clusters by
noting that
after updating the cluster assignment of each observation, the
parameters of any unused empty clusters are in fact already
independently and
identically distributed according to the base distribution. Thus,
we can consider reusing them for updating
the next observation. However, note that as a result the parameters of
the empty clusters used in different updates will not be independent,
and the justification of correctness of Neal's Algorithm 8 (as Gibbs
sampling in an augmentation scheme) is no longer valid.

In this section we develop an algorithm that does reuse new clusters,
and show using a different technique that it is valid with stationary
distribution given by the posterior. For the new algorithm, we instead
augment the MCMC state space with a permanent set of $C$ empty
clusters, so the augmented state space now consists of the partition
$\bfpi$, the latent variable $U$, the parameters $\{X^\ast_c \dvtx c\in
\bfpi\}$ of existing clusters and the parameters $\{X^\mathrm{e}_k \dvtx k\in[C]\}$ of the auxiliary empty clusters. Further, we develop the
cluster assignment updates as Metropolis--Hastings updates instead of
Gibbs updates.

In the following we use the superscript $'$ in order to denote
variables and values associated with the new proposed state of the
Markov chain.
Suppose we wish to update the cluster assignment of observation $Y_i$.
Again we introduce the variable $z_i$, which takes value $c\in\bfpi
_{\setminus i}$ if $Y_i$ is assigned to a cluster containing other
observations, and takes values $k\in[C]$ uniformly at random if $Y_i$
is assigned to a cluster by itself.
If $z_i=c\in\bfpi_{\setminus i}$, then the proposal distribution
$\mathbb{Q}$ is described by a two-step algorithm:
\begin{enumerate}[2b.]
\item[1.] Sample the variable $z_i'$ from the conditional distribution
\eqref{eq:post_zi_2} as before.
\item[2a.] If $z_i'=c'\in\bfpi_{\setminus i}$, then we simply assign
$Y_i$ to the existing cluster $c'$.
\item[2b.]
If $z_i'=k'$ for one of the empty clusters $k'\in[C]$ with $X^\mathrm
{e}_{k'}=x$, then:
\begin{enumerate}[(ii)]
\item[(i)] we assign $Y_i$ to a newly created cluster with parameter
$X^{\ast\prime}_\varnothing :=x$;
\item[(ii)] set $X^{\mathrm{e}\prime}_{k'} := x'\sim\mu_0$ with a new draw
from the base distribution.
\end{enumerate}
\end{enumerate}
On the other hand, if $Y_i$ is currently assigned to a cluster all by
itself, say, with parameter $X^\ast_\varnothing=x_0$, then $z_i$ will
initially take on each value $k\in[C]$ uniformly with probability
$1/C$. We start by setting the value for the $k$th empty cluster
parameter $X^\mathrm{e}_k:=x_0$ (its old value is discarded) and then
removing the singleton cluster that $Y_i$ is currently assigned to.
Then the two-step algorithm above is carried out.

It is important to point out that the proposal described above is
reversible. For example, the reverse of moving $Y_i$ from an existing
cluster $c$ to a new cluster with parameter $X^\ast_\varnothing=x$, where
$x$ is the previous value of $X^\mathrm{e}_{k'}$ with its new value being
a draw $x'$ from $\mu_0$, is exactly the reverse of the proposal moving
$Y_i$ from a singleton cluster with parameter $X^\ast_\varnothing=x$ to
the cluster~$c$, while replacing the previous value $x'$ of $X^\mathrm
{e}_{k'}$ with $x$. We denote the two proposals as $(c\Rightarrow k')$
and $(k'\Rightarrow c)$. Analogously, the reverse of $(c\Rightarrow
c')$ is $(c'\Rightarrow c)$ and the reverse of $(k\Rightarrow k')$ is
$(k'\Rightarrow k)$.

Note also that the proposals are trans-dimensional since the number of
clusters in the partition $\bfpi$ (and particularly the number of
cluster parameters) can change. See Green~\cite{Gre1995a} and
Richardson and Green~\cite{RicGre1997a} for approaches to
trans-dimensional MCMC. Fortunately they are dimensionally balanced. In
fact, we can show that the acceptance probability is simply always one.
For example, for the $(c\Rightarrow k')$ proposal, the joint
probability of the initial state and the proposal probability are,
respectively, proportional to
\[
\mathbb{P}\bigl[ z_i = c,X^\mathrm{e}_{k'} \in
dx\cdots\bigr]\propto\frac
{\kappa
_{|c|+1}(U)}{\kappa_{|c|}(U)}f\bigl(Y_i |
X^\ast_c\bigr) \mu_0(dx)
\]
and
\begin{eqnarray*}
&&\mathbb{Q}\bigl[z_i' = k',X^{\ast\prime}_\varnothing
\in dx_0,X^{\mathrm
{e}\prime}_{k'} \in dx'\cdots |\\
&&\hspace*{63pt}\quad z_i = c,X^\mathrm{e}_{k'} \in dx\cdots\bigr]\\
&&\quad\propto
\frac{\kappa_1(U)}{C} f(Y_i | x) \delta_x(dx_0)
\mu_0\bigl(dx'\bigr).
\end{eqnarray*}
We have suppressed listing all other variables for brevity.
For the reverse proposal $(k'\Rightarrow c)$, the probabilities are,
respectively,
\begin{eqnarray*}
&&\mathbb{P}\bigl[ z_i' = k',dX^{\ast\prime}_{\varnothing}
\in dx_0,X^{\mathrm
{e}\prime}_{k'} \in dx'\cdots
\bigr]\\
&&\quad\propto \frac{1}{C} \kappa_1(U) \mu_0(dx_0)f(Y_i
| x)\mu_0\bigl(dx'\bigr)
\end{eqnarray*}
and
\begin{eqnarray*}
&&\mathbb{Q}\bigl[z_i = c,X^\mathrm{e}_{k'} \in
dx\cdots | z_i' = k',X^{\ast\prime}_{\varnothing}
\in dx_0\cdots\bigr]\\
&&\quad\propto\frac{\kappa
_{|c|+1}(U)}{\kappa_{|c|}(U)}f\bigl(Y_i |
X^\ast_c\bigr)\delta_{x_0}(dx).
\end{eqnarray*}
Note that the normalization constants arising from the conditional
distributions \eqref{eq:post_zi_2} for proposals in both directions are
the same, so they can be ignored. We see that the product of the
probabilities for the $(c\Rightarrow k')$ proposal is the same as that
for the reverse $(k'\Rightarrow c)$, so the Metropolis--Hastings
acceptance ratio is simply one. Similarly, the acceptance ratios of
other proposal pairs are also equal to one.


In addition to updating the cluster assignments of all observations as
above, we also need to update the parameters of the $C$ empty clusters.
We do this by marginalizing them out before updating $U$ and the
hyperparameters according to Section~\ref{sec:hyperparam}, and
replacing them afterward with new independent and identically
distributed draws from the base distribution. Note that the resulting
Metropolis--Hastings updates are very similar to the augmentation scheme
Gibbs updates described in Section~\ref{sec:algorithm8}. The only
difference is the way the parameters of the empty clusters are managed
and retained across cluster assignment updates of multiple observations.

\subsection{Conditional Slice Sampler}\label{sec:slice}

In the so-called marginalized samplers the CRM $\mu$ is marginalized
out while the latent variables $\mathbf{X}$ representing the partition
structure and the cluster parameters are sampled.
In a conditional sampler we instead alternatively Gibbs sample $\mu$
given $\mathbf{X}$ and $\mathbf{X}$ given $\mu$. Proposition \ref
{th_posterior} provides the conditional distribution for ${\mu}$ given
$\mathbf{X}$, while the conditional of $\mathbf{X}$ given ${\mu}$ is
straightforward. What is not straightforward is the fact that since
${\mu}$ has an infinite number of atoms we cannot explicitly sample all
of it on a computer with finite resources. Thus, it is necessary to
truncate $\mu$ and work only with a finite number of atoms.

In this section we will describe a conditional sampler based on a slice
sampling strategy for truncation. See Walker~\cite{Wal2007a} for the
slice sampler in DP mixture models, and Griffin and Walker~\cite
{GriWal2011a} and Griffin et al.~\cite{GriKolSte2011a} for slice
samplers in NRM mixture models on which our sampler is based.
Recall from \eqref{eq:nrmmixture} that each observation $Y_i$ is
assigned to a cluster parametrized by an atom $X_i$ of $\mu$. We
augment the state with an additional slice variable $S_i$, whose
conditional distribution is a Uniform distribution taking values
between 0 and the mass of atom $X_i$ in $\mu$, that is,
%
%
\begin{equation}
S_i | X_i,\mu\sim\Unif \bigl(0,\mu\bigl(
\{X_i\}\bigr) \bigr). \label{eq:slice}
\end{equation}
Marginalizing out $S_i$, the joint distribution of the other variables
reduces back to the desired posterior distribution. On the other hand,
conditioned on $S_i$, $X_i$ can only take on values corresponding to
atoms in $\mu$ with mass at least $S_i$. Since $S_i$ is almost surely
positive, this set of atoms is finite, and so $S_i$ effectively serves
as a truncation level for $\mu$ in the sense that only these finitely
many atoms are needed when updating $X_i$. Over the whole data set,
only the (finitely many) atoms in $\mu$ with mass at least
$
S=\min_{i\in[n]} S_i>0
$
are required when updating the set of latent variables $\mathbf{X}$
given $\mu$ and the slice variables $\mathbf{S} = \{S_i \dvtx i\in[n]\}$.

The state space of our sampler thus consists of the latent variables
$\mathbf{X}$, the slice variables $\mathbf{S}$, the CRM $\mu$ and the
auxiliary variable $U$ introduced in Section~\ref{sec:nrmmixture}.
At a high level, our sampler is simply a Gibbs sampler, iterating among
updates to $\mathbf{X}$, $U$, and both $\mu$ and $\mathbf{S}$ jointly.

First consider updating $\mathbf{X}$. It is easy to see that
conditioned on $U$, $\mu$ and $\mathbf{S}$ the $X_i$'s are mutually
independent. For each $i\in[n]$, the conditional probability of $X_i$
taking on value $x\in\mathbb{X}$ is proportional to the product of the
probability $ \mu(\{x\})/\mu(\mathbb{X})$ of $x$ under the NRM
$\tilde
{\mu}$, the conditional distribution function $f(Y_i | x)$ of
observation $Y_i$, and the conditional density of $S_i$ given $X_i=x$,
which is simply $1/\mu(\{x\})$ when $0<S_i<\mu(\{x\})$ and 0 otherwise.
The resulting conditional distribution of $X_i$ simplifies to
\[
\mathbb{P} [X_i=x | \mu,Y_i,S_i ] \propto
\cases{ f(Y_i | x), &$\mbox{if $S_i<\mu
\bigl(\{x\}\bigr)$,}$\vspace*{2pt}
\cr
0, &$\mbox{otherwise}.$} %
\]
This is a discrete distribution, with positive probability of $X_i=x$
only when $x$ coincides with the location of an atom in $\mu$ with mass
greater than $S_i$. Note that there almost surely are only a finite
number of such atoms in $\mu$ since $S_i>0$, so that updating $X_i$ is
computationally feasible.

Now consider updating $U$. We will perform this update conditioned only
on the partition described by $\mathbf{X}$, with the random measure
$\mu
$ and the slice variables $\mathbf{S}$ marginalized out. We can also
update any hyperparameters of the CRM and of the base distribution $\mu
_0$ at this step as well. For example, if $\tilde\mu$ is a NGGP, we can
update both $U$ and the parameters $(a,\sigma,\tau)$ using those
described in Section~\ref{sec:hyperparam}, which makes these
Metropolis-within-Gibbs updates.

Finally, consider updating $\mu$ and $\mathbf{S}$ jointly. Note that
this update needs to be performed right after the $U$ update since $\mu
$ and $\mathbf{S}$ were marginalized out when updating $U$. The
conditional distribution of $\mu$ given $U$ and $\mathbf{X}$ is given
by Proposition~\ref{th_posterior}, which shows that $\mu$ will contain
a finite number of fixed atoms located at the unique values $\{X^\ast_c
\dvtx c\in\boldsymbol{\pi}\}$ among $\mathbf{X}$, and a countably infinite
number of randomly located atoms corresponding to the unused clusters
in the NRM mixture model. Given $\mu$, the slice variables are
independent with distributions given by \eqref{eq:slice}; in
particular, note that they depend only on the masses of the fixed atoms
of $\mu$. On the other hand, as noted above, we only need the random
atoms of $\mu$ with masses above the overall truncation level $S=\min_{i\in[n]} S_i$. Therefore, a sufficient method for sampling both $\mu$
and $\mathbf{S}$ is to first sample the fixed atoms of $\mu$, followed
by $\mathbf{S}$, and finally the random atoms with masses above $S$.

For the fixed atoms of $\mu$, Proposition~\ref{th_posterior} states
that each of them corresponds to a unique value among $\mathbf{X}$ and
that their masses are mutually independent and independent from the
random atoms. For each such unique value $X^\ast_c$, $c\in\bfpi$, the
conditional distribution of its mass $J'_c$ is
%
%
\begin{equation}
\label{eq:probj} \mathbb{P} \bigl[J'_c\in ds | U,
\mathbf{X} \bigr] \propto s^{|c|}e^{-Us} \rho(ds),
\end{equation}
where $|c|$ is the number of observations allocated to the cluster $c$,
that is, with $X_i=X^\ast_c$. Under the assumption of $\tilde{\mu}$
being a NGGP, the density in \eqref{eq:probj} simplifies to
$s^{|c|-\sigma-1}e^{-(U+\tau)s}$, a Gamma density. We also update the
locations of the fixed atoms as well using an acceleration step as in
Bush and MacEachern~\cite{BusMac1996a}. The conditional distribution
function of $X_c^\ast$ is proportional to its prior distribution
function times the likelihoods of observations assigned to the cluster,
that is,
\[
\mathbb{P} \bigl[X_c^\ast\in dx | \mathbf{Y} \bigr]
\propto\mu _0(dx)\prod_{i\in c}
f(Y_i | x),
\]
where $i\in c$ indicates indices of those observations assigned to the
cluster $c$. Note that any ergodic Markov kernel with the above as its
stationary distribution suffices.

Once the fixed atoms are updated, the slice variables are updated by
sampling each $S_{i}$ independently from its conditional distribution
\eqref{eq:slice}.
Finally, the random atoms of $\mu$ with mass above the overall
truncation level $S$ can be sampled using Proposition \ref
{th_posterior}. As we work only with homogeneous CRMs here, the
locations are simply independent and identically distributed draws from
$\mu_0$, while their masses are distributed according to a Poisson
random measure on $[S,\infty)$ with an exponentially tilted intensity
measure $\rho'(ds)=e^{-Us}\rho(ds)$.

\begin{figure}

\includegraphics{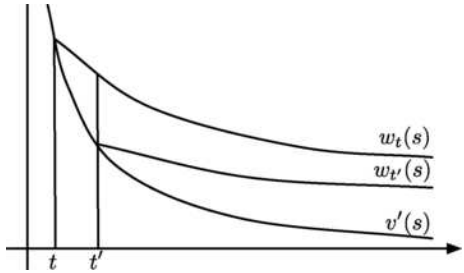}

\caption{Adaptive bounds for simulating from a Poisson random measure
with intensity $v'(s)$.}\label{fig:adaptivethinning}
\end{figure}

We propose an adaptive thinning approach (see Ogata~\cite{Oga1981a}) to
sample from the Poisson random measure which is computationally
efficient but applies only to certain classes of intensity measures
which can be adaptively bounded in the following sense.
Let $v'(s)$ be the density of $\rho'(ds)$ with respect to the Lebesgue
measure and assume that for each $t\in\mathbb{R}_+$ there is a function
$w_t(s)$ such that $w_t(t)=v'(t)$ and $w_t(s)\ge w_{t'}(s)\ge v'(s)$
for every $s,t'\ge t$. See Figure~\ref{fig:adaptivethinning}. In
particular, for the NGGP one has
\[
v'(s) = \frac{a}{\Gamma(1-\sigma)}s^{-1-\sigma}e^{-s(\tau+U)},
\]
and we can use the family of adaptive bounds
\[
w_t(s) = \frac{a}{\Gamma(1-\sigma)}t^{-1-\sigma}e^{-s(\tau+U)},
\]
with the inverse of the integral given by
\[
W^{-1}_t(r)=t-\frac{1}{\tau+U}\log \biggl(1-
\frac{r(\tau+U)\Gamma
(1-\sigma
)}{at^{-1-\sigma}e^{-t(\tau+U)}} \biggr).
\]
Note that both $w_t(s)$ and the inverse of the map $W_t(s)= \int_t^s
w_t(s')\,ds'$ are analytically tractable, with $\int_t^\infty
w_t(s')\,ds'<\infty$.

%

The method is based on the idea of thinning by Lewis and Shedler~\cite
{LewShe1979a}, a method to simulate from a Poisson random measure by
first proposing points according to a proposal Poisson random measure
with higher intensity than the desired one. Each point is then accepted
with probability given by the ratio of intensities under the proposal
and desired Poisson random measures. The idea of adaptive thinning is
that we can propose points iteratively from left to right starting at
$S$, and after each proposed point $t$ the bound $w_{t}$ is used as the
intensity of the proposed Poisson random measure from which the next
point is drawn. As $t$ increases, the bound tightens, so rejections are
reduced. Further, as $\int_s^\infty w_t(s')\,ds'<\infty$, the iteration
will terminate after a finite number of points are proposed.
Specifically, the sampling scheme is described as follows:
\begin{enumerate}[2.]
\item[1.] set $N:=\varnothing$, $t := S$;
\item[2.] iterate until termination:
\begin{enumerate}[(iii)]
\item[(i)] let $r$ be a draw from an Exponential distribution with
parameter $1$;
\item[(ii)] if $r>W_t(\infty)$, terminate; else set $t':= W^{-1}_t(r)$;
\item[(iii)] with probability $v'(t')/w_t(t')$ accept sample: set
$N:=N\cup\{t'\}$;
\item[(iv)] set $t:=t'$ and continue to next iteration;
\end{enumerate}
\item[3.] return $N$ as a draw from the Poisson random measure with
intensity $v'$ on $[S,\infty)$.
\end{enumerate}
The returned $N$ constitutes the set of masses for the random atoms in
$\mu$ with masses above the overall truncation level $S$.


\subsection{Some Remarks}

There is a rich literature on conditional sampling schemes for
nonparametric mixture models.
In the DP mixture model case, the use of the stick-breaking
representation for $\tilde{\mu}$, as proposed by Ishwaran and James~\cite{IshJam2001a}, Papaspiliopoulos and Roberts~\cite{PapRob2008a} and
Walker~\cite{Wal2007a}, is very simple since it involves a sequence of
random variables that are independently Beta distributed a priori as
well as a posteriori conditioned on other variables. However, this
simplicity comes at a cost of slower mixing due to the label-switching
problem discussed in Jasra et al.~\cite{JasHolSte2005a}.
Papaspiliopoulos and Roberts~\cite{PapRob2008a} noted that while the
likelihood is invariant to the ordering of atoms, the stick-breaking
prior has a weak preference for atoms to be sorted by decreasing mass,
resulting in multiple modes in the posterior. Then, they proposed
Metropolis--Hastings moves that interchange pairs of atoms to improve
mixing. A more sophisticated approach that avoids the weak
identifiability altogether is to use the natural unordered
representation stated in Proposition~\ref{th_posterior}. This approach
was taken in Griffin and Walker~\cite{GriWal2011a}, and we used it here
as well.

There are a few alternative methods for sampling from the Poisson
random measure governing the\break masses of the random atoms. Griffin and
Walker~\cite{GriWal2011a} proposed first sampling the number of atoms
from a Poisson with rate $\rho'([S,\infty))$, then sampling the masses
independently and identically distributed
according to a distribution obtained by normalizing~$\rho'$. Another
possibility proposed by Barrios et al.~\cite{BarLijNie2012a} and
Nieto-Barajas and Pr\"unster~\cite{NiePru2009a} is to use the
representation proposed by Ferguson and Klass~\cite{FerKla1972a}, which
involves using the mapping theorem for Poisson random measures to
sample the masses in order starting from the largest to the smallest.


Our slice sampler follows Griffin and Walker~\cite{GriWal2011a} in
introducing a slice variable $S_i$ for each observation $i$. Another
approach described in Griffin and Walker is to introduce a single slice
variable $S_\mathrm{all}$ for all observations, with conditional distribution
\[
S_\mathrm{all} \sim\Unif \Bigl( 0,\min_{i\in[n]} \mu\bigl(
\{X_i\} \bigr) \Bigr).
\]
Griffin and Walker~\cite{GriWal2011a} found that either method may work
better than the other in different situations. We preferred the method
described here, as it is simpler and the updates for the latent
variables, which form the most time consuming part of the algorithm,
can be trivially parallelized to take advantage of recent parallel
computation hardware architectures.

Slice samplers have the advantages that they can technically be exact
in the sense that they target the true posterior distribution. This is
opposed to alternative truncations which introduce approximations by
ignoring atoms with low masses, for example, Ishwaran and James~\cite
{IshJam2001a} and Barrios et al.~\cite{BarLijNie2012a}.
However, a difficulty with slice samplers is that although the number
of random atoms in $\mu$ above the truncation level $S$ is finite with
probability one, the actual number generated can occasionally be
extremely large, for example, in case of NGGPs when $S$ is small and
$\sigma$ is large. In our implementation our program can occasionally
terminate as it runs out of memory. We fix this by introducing an
approximation where we only generate atoms with masses above $10^{-8}$
and only keep a maximum of the $10^6$ atoms with largest masses.
Griffin and Walker~\cite{GriWal2011a} and Barrios et al.~\cite
{BarLijNie2012a} have also made similar approximations. Of course this
approximation effectively nullifies the advantage of slice samplers
being exact, though we have found in experiments that the
approximation introduced is minimal.

Comparing the computational requirements of the proposed marginalized
and conditional samplers, we expect the marginalized samplers to
produce chains with less autocorrelation since they marginalize more
latent variables out. Further, their computational costs per iteration
are controllable and more stable since each involves introducing a
fixed number of empty clusters. Concluding, while the conditional
sampler is easily parallelizable, the marginalized samplers are not.


\section{Numerical Illustrations}\label{sec4}

In this section we illustrate the algorithms on a number of data sets:
three simple and well-studied data sets, the galaxy, acidity and the
Old Faithful geyser data sets, as well as a more complex data set of
neuronal spike waveforms. The galaxy data set consists of the
velocities at which 82 galaxies are receding away from our own and the
acidity data set consists of the log acidity measurements of 155 lakes
in Wisconsin; both are one-dimensional. The geyser data set is
two-dimensional, consisting of 272 durations of eruptions along with
the waiting times since the last one. The spikes data set\setcounter{footnote}{1}\footnote{We
thank G\"or\"ur and Rasmussen~\cite{GorRas2004a} for providing us with
the data set.} consists of a total of 14,802 neuronal spike waveforms
recorded using tetrodes. Each of the four electrodes contributes 28
readings sampled at 32~kHz, so that each waveform is 112-dimensional.
Prototypical waveforms are shown in Figure~\ref{fig:spikes}. To reduce
computation time, in the following we first used PCA to reduce the data
set down to six dimensions, which preserved approximately 80\% of the
variance and sufficient information for the mixture model to recover
distinct clusters.

%
\begin{figure*}

\includegraphics{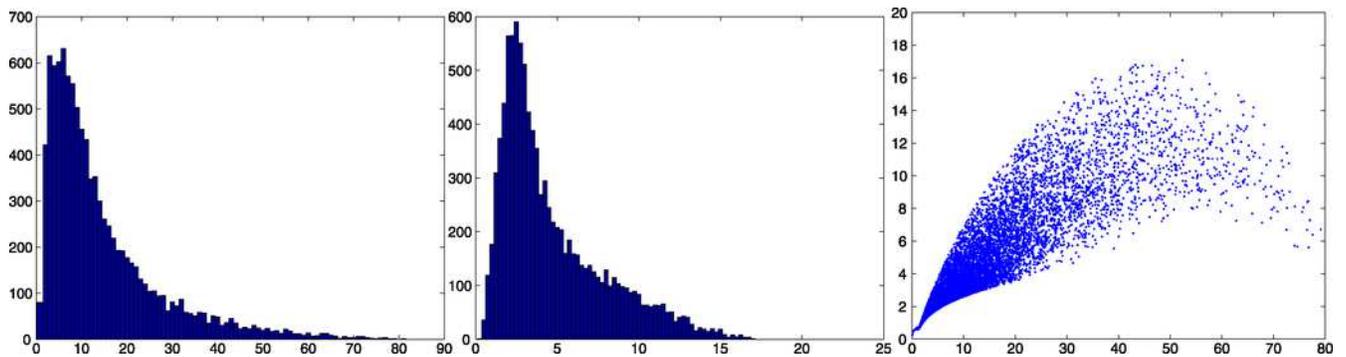}

\caption{Visualizing the induced prior on the number of clusters with
$n=82$ corresponding to the size of the galaxy data set. Left:
histogram of the mean number of clusters. Center: histogram of the
standard deviation of the number of clusters. Right: scatter plot of
standard deviation vs mean of the number of clusters. 10,000 draws from
the prior for $a$ and $\sigma$ were used.}
\label{fig:clusterprior}
\end{figure*}

We analyzed the data sets by means of NRM mixtures of (multivariate)
Gaussian distributions. Let $D$ be the number of dimensions of the data
set. The base distribution over the Gaussian means and covariance
matrices is factorized as follows:
\[
\mu_0(dm,d\Sigma) =\mathcal{N}_D(dm;m_0,S_0)
\mathcal{IW}_D(d\Sigma ;\alpha_0,\Sigma_0),
\]
where $\mathcal{N}_D$ denotes a $D$-dimensional Gaussian distribution
with given mean and covariance matrix and $\mathcal{IW}_D$ denotes an
inverse Wishart over $D\times D$ positive definite matrices with given
degree of freedom and scale matrix.

A number of authors have advocated the use of weakly informative priors
for mixtures of Gaussian distributions. See Nobile~\cite{Nob1994a},
Raftery~\cite{Raf1996a} and Rich\-ardson and Green~\cite{RicGre1997a}. We
follow the approach advocated by Richardson and Green~\cite
{RicGre1997a}, generalizing it to the multivariate setting. In
particular, we assume knowledge of a likely range over which the data
lies, with the range in the $i$th dimension being $[m_{0i}-s_i,
m_{0i}+s_i]$. We set $S_0$ to be a diagonal matrix with $i$th diagonal
entry being $s_i^2$ so that the prior over component means is rather
flat over the range. We set $\alpha_0=D+3$, and set a hierarchical
prior $\Sigma_0\sim\mathcal{IW}_D(\beta_{0},\gamma_{0} S_0)$ where
$\beta_{0}$ is chosen to be $D-0.6$. These degrees of freedom express
the prior belief that component covariances are generally similar
without being informative about their absolute scales. We choose
$\gamma
_{0}$ so that $\mathbb{E}[\Sigma] = S_0/50$, that is, that the a
priori range of each component is approximately $\sqrt{50}\approx7$
times smaller than the range set by $S_0$, although the model is not
sensitive to this prior range since $\Sigma_0$ is random and allowed to
adapt to the data in its posterior. In the one-dimensional setting this
prior reduces to the same one used by Richardson and Green. A detailed
study of prior specifications for mixtures of multivariate Gaussian
distributions is beyond the scope of this paper and the interested
reader is referred to M\"uller et al.~\cite{MulErkWes1996a} and Fraley
and Raftery~\cite{FraRaf2007a} for alternative specifications.

In the one-dimensional setting we also considered a conjugate prior so
that we can compare the samplers with and without component parameters
mar\-ginalized out. We use a similar weakly informative prior in
the\vadjust{\goodbreak}
conjugate case as well, with base distribution given by
\begin{eqnarray*}
&&\mu_0(dm,d\Sigma) \\
&&\quad= \mathcal{N}_1\bigl(dm;m_0,S_0
\Sigma_0^{-1}\Sigma \bigr)\mathcal {IW}_1(d
\Sigma;\alpha_0,\Sigma_0),
\end{eqnarray*}
where the one-dimensional inverse Wishart with parameter $(a,s)$ is
simply an inverse gamma with parameter $(a/2,s/2)$.
We used the same $\alpha_0$ and hierarchical prior for $\Sigma_0$ as
for the nonconjugate prior, while the a priori expected value for the
variance of $m$ can be seen to be $\mathbb{E}[S_0\Sigma_0^{-1}\Sigma
]=S_0$, which is independent of $\Sigma_0$ and matches the nonconjugate
case. In both cases we updated the $\Sigma_0$ by Gibbs sampling.

The parameters $a$ and $\tau$ of the NGGP are redundant (see
Section~\ref{sec:nrm} for details), so we simply set $\tau=1$ in the
simulations. We place a gamma $(1,1)$ prior on $a$, while~$\sigma$ is
given a beta prior with parameters $(1,2)$. We can visualize the
induced prior on the partition structure by drawing samples of $a$ and
$\sigma$ from their prior and for each sample calculating the mean and
standard deviation of the prior over the number of clusters.
Figure~\ref{fig:clusterprior} shows the result for $n=82$, corresponding to the
size of the galaxy data set. We see that the prior gives support over a
wide range of values for the mean and standard deviation of the number
of clusters, with higher probability for the mean number of clusters to
be in the region between 1 and 20.

%
\begin{figure*}

\includegraphics{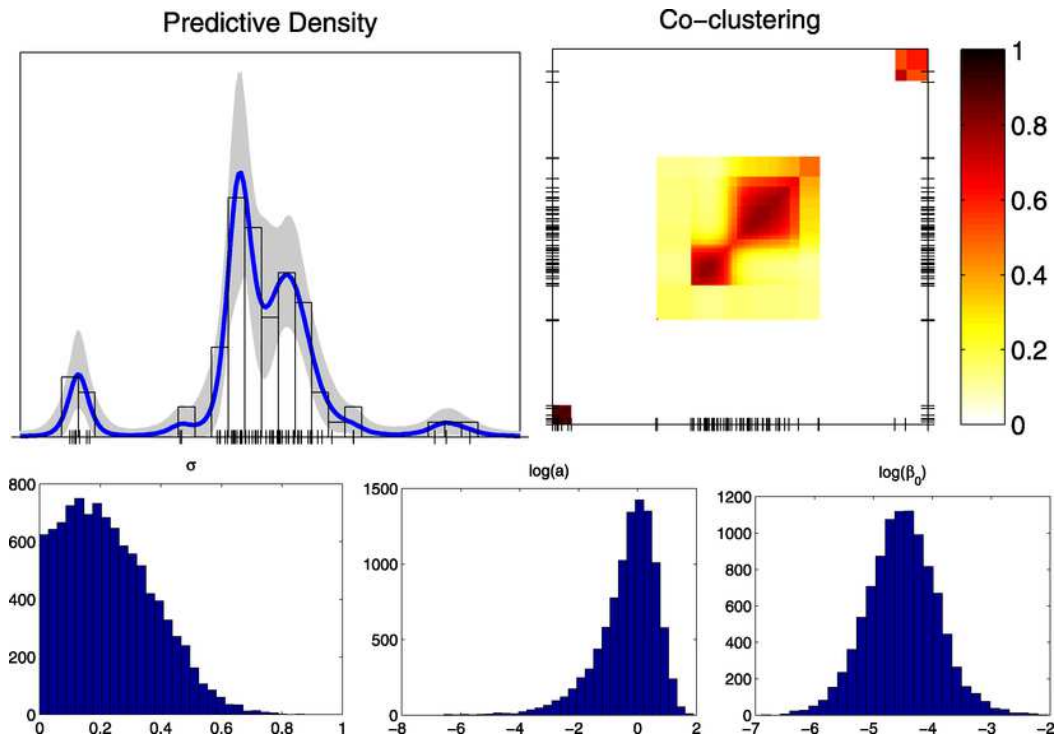}

\caption{Visualizations of the posterior distribution of the
nonconjugate NGGP mixture model on the galaxy data set. Top-left:
posterior mean and 95\% credible interval (pointwise) of the density
function. Top-right: co-clustering probabilities, whiskers at edges
denote observations. Bottom: histograms of the posteriors of $\sigma$,
$\log(a)$ and $\log(\beta_0)$, respectively.}\label{fig:galaxy}
\end{figure*}

\subsection{One-Dimensional Data Sets: Galaxy and Acidity}
In the conjugate case, we applied both the conjugate marginalized
sampler of Section~\ref{sec:conjmarg} and the conditional slice sampler
of Section~\ref{sec:slice} (but with mixture component parameters
marginalized out). To investigate the difference between marginalizing
out the component parameters and not, we also applied the
generalization of Neal's Algorithm 8 in Section~\ref{sec:nonconjmarg}
and the Reuse algorithm of Section~\ref{sec:reuse}, both with $C\in\{
1,2,3,4,5\}$ and the conditional slice sampler to the conjugate model
(sampling the parameters instead of marginalizing them out). In the
nonconjugate case we applied the conditional slice sampler and the two
nonconjugate marginalized samplers with $C\in\{1,2,3,4,5\}$. For all
samplers in both conjugate and nonconjugate models, the initial\break 10,000
iterations were discarded as burn-in, followed by 200,000 iterations,
from which we collected 10,000 samples.

%
\begin{figure*}[t!]

\includegraphics{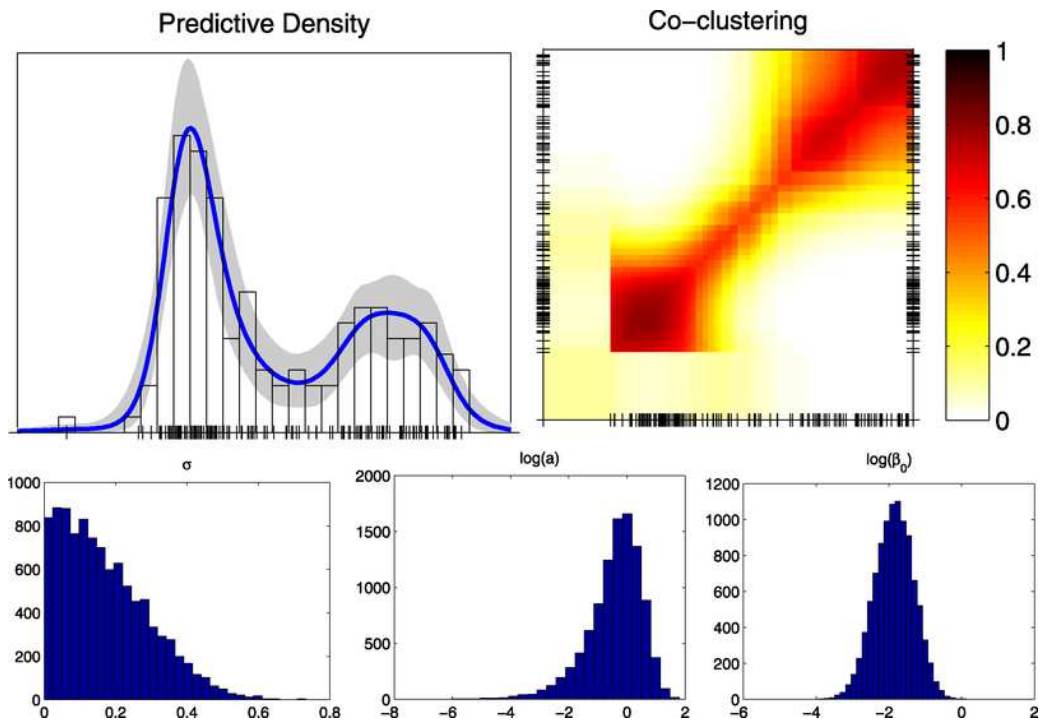}

\caption{Visualizations of the posterior distribution of the
nonconjugate NGGP mixture model on the acidity data set. Top-left:
posterior mean and 95\% credible interval (pointwise) of the density
function. Top-right: co-clustering probabilities, whiskers at edges
denote observations. Bottom row: histograms of the posterior of $\sigma
$, $\log(a)$ and $\log(\beta_0)$, respectively.}\label{fig:acid}
\end{figure*}

\begin{figure*}[b!]

\includegraphics{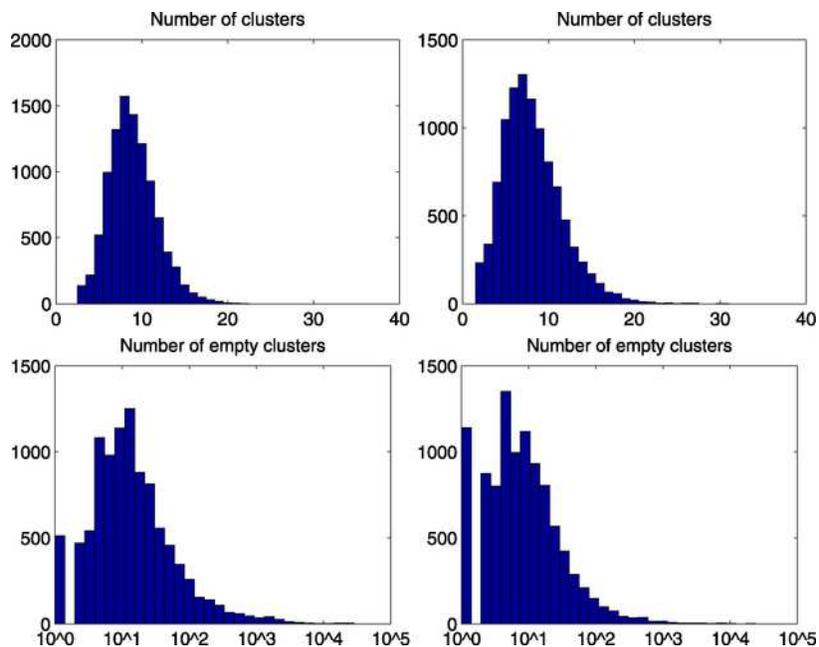}

\caption{Top: Distribution of the number of components used,
for the nonconjugate NGGP mixture model
for the galaxy (left) and the acidity (right) data sets, respectively.
Bottom: distribution of the number of empty clusters instantiated by
the conditional slice sampler at each iteration (on the logarithmic
scale) for the galaxy (left) and the acidity (right) data sets.}\label
{fig:numclusters}
\end{figure*}
%
\begin{table*}
\tabcolsep=0pt
\caption{Comparison of sampler efficiencies on the one-dimensional
galaxy and acidity data sets. Each of 10 runs produces 10,000 samples,
at intervals of 20 iterations, after an initial burn-in period of 10,000
iterations. Each entry reports the average and standard error over the
10 runs. In the first column, C indicates conjugate prior
specification, N for nonconjugate, while M indicates component
parameters are marginalized and S means they are sampled}\label
{table:efficiencies}
\begin{tabular*}{\textwidth}{@{\extracolsep{\fill}}lccccc@{}}
\hline
&  &\multicolumn{2}{c}{\textbf{Galaxy}} & \multicolumn
{2}{c@{}}{\textbf{Acidity}} \\
\ccline{3-4,5-6}
\textbf{Model}& \textbf{Sampler}& \textbf{Runtime (s)} & \textbf{ESS} & \textbf{Runtime (s)} & \textbf{ESS} \\
\hline
CM & Cond Slice&
$239.1\pm4.2$ & $2004\pm178$ &$196.5\pm1.0$ & $ 910\pm142$ \\
CM & Marg ($C=1$) &
$215.7\pm1.4$ & $7809\pm87$ &$395.5\pm1.7$ & $5236\pm181$ \\[6pt]
CS & Cond Slice&
$133.0\pm3.2$ & $1594\pm117$ &$77.4\pm0.7$ & $1099\pm49$ \\
CS & Marg Neal 8 ($C=1$) &
$74.4\pm0.6$ & $5815\pm145$ &$133.3\pm1.8$ & $4175\pm85$ \\
CS & Marg Neal 8 ($C=2$) &
$87.9\pm0.6$ & $6292\pm94$ &$163.8\pm1.5$ & $4052\pm158$ \\
CS & Marg Neal 8 ($C=3$) &
$101.9\pm0.7$ & $6320\pm137$ &$188.2\pm1.1$ & $4241\pm99$ \\
CS & Marg Neal 8 ($C=4$) &
$115.9\pm0.6$ & $6283\pm86$ &$216.6\pm1.7$ & $4266\pm122$ \\
CS & Marg Neal 8 ($C=5$) &
$130.0\pm0.6$ & $6491\pm203$ &$243.8\pm2.0$ & $4453\pm123$ \\
CS & Marg Reuse ($C=1$) &
$64.3\pm0.3$ & $4451\pm79$ &$114.6\pm2.0$ & $3751\pm65$ \\
CS & Marg Reuse ($C=2$) &
$67.6\pm0.5$ & $5554\pm112$ &$123.1\pm1.9$ & $4475\pm110$ \\
CS & Marg Reuse ($C=3$) &
$71.3\pm0.5$ & $5922\pm157$ &$128.2\pm2.2$ & $4439\pm158$ \\
CS & Marg Reuse ($C=4$) &
$74.9\pm0.5$ & $6001\pm101$ &$140.1\pm1.6$ & $4543\pm108$ \\
CS & Marg Reuse ($C=5$) &
$78.7\pm0.6$ & $6131\pm124$ &$147.7\pm1.5$ & $4585\pm116$ \\[6pt]
NS & Cond Slice&
$75.5\pm1.2$ & $ 939\pm92$ &$50.9\pm0.5$ & $ 949\pm70$ \\
NS & Marg Neal 8 ($C=1$) &
$65.0\pm0.5$ & $4313\pm172$ &$110.9\pm0.8$ & $4144\pm64$ \\
NS & Marg Neal 8 ($C=2$) &
$78.6\pm0.4$ & $4831\pm168$ &$139.2\pm1.8$ & $4290\pm125$ \\
NS & Marg Neal 8 ($C=3$) &
$92.5\pm0.5$ & $4785\pm97$ &$162.7\pm0.9$ & $4368\pm72$ \\
NS & Marg Neal 8 ($C=4$) &
$106.3\pm0.5$ & $4849\pm120$ &$187.6\pm1.1$ & $4234\pm142$ \\
NS & Marg Neal 8 ($C=5$) &
$119.7\pm0.6$ & $5029\pm89$ &$215.4\pm1.3$ & $4144\pm213$ \\
NS & Marg Reuse ($C=1$) &
$55.2\pm0.5$ & $3830\pm103$ &$91.3\pm0.9$ & $4007\pm122$ \\
NS & Marg Reuse ($C=2$) &
$58.7\pm0.5$ & $4286\pm101$ &$98.1\pm0.9$ & $4192\pm138$ \\
NS & Marg Reuse ($C=3$) &
$62.4\pm0.6$ & $4478\pm124$ &$105.1\pm0.9$ & $4260\pm136$ \\
NS & Marg Reuse ($C=4$) &
$66.1\pm0.5$ & $4825\pm63$ &$112.3\pm1.0$ & $4191\pm139$ \\
NS & Marg Reuse ($C=5$) &
$69.8\pm0.6$ & $4755\pm141$ &$121.0\pm1.8$ & $4186\pm121$ \\
\hline
\end{tabular*}
\end{table*}

Figure~\ref{fig:galaxy} shows some aspects of the posterior
distribution on the galaxy data set for the nonconjugate model obtained
using the conditional slice sampler, while Figure~\ref{fig:acid} shows
the same for the acidity data set. The marginalized samplers produce
the same results, while the posterior for the conjugate model is
similar and not shown.
The co-clustering probabilities are computed as follows: the color at
location $(x,y)$ indicates the posterior probability that observations
$Y_i$ and $Y_j$ belong to the same components, where $Y_i$ is the
largest observed value smaller than $\min(x,y)$ and $Y_j$ is the
smallest observed value larger than $\max(x,y)$. The posterior
distribution of the number of components used is shown in the top half
of Figure~\ref{fig:numclusters}. The posterior distributions are
consistent with those obtained by previous authors, for example,
Richardson and Green~\cite{RicGre1997a}, Escobar and West~\cite
{EscWes1995a}, Griffin and Walker~\cite{GriWal2011a} and Roeder~\cite
{Roe1994a}.

In Table~\ref{table:efficiencies} we compared the samplers in terms of
both their run times (in seconds, excluding time required to compute
predictive probabilities) and their effective sample sizes (ESSs) of
the number of components (as computed using the R Coda package). By
marginalizing out the mixture component parameters, we see that the
samplers mix more effectively with higher ESSs. The conditional slice
sampler and the nonconjugate marginalized samplers were effective at
handling mixture component parameters that were sampled instead of
marginalized out, but the ESSs were a little lower, as expected. Among
the marginalized samplers, with increasing $C$ both the computational
costs and the ESS generally increase, with the computational cost of
Neal's Algorithm 8 increasing more rapidly, as expected.

While the conditional slice sampler is typically faster than the
marginalized samplers, they also produce lower ESSs. An important
difference between the slice sampler and the marginalized samplers is
that the slice sampler we proposed uses one slice variable per
observation, so typically a significantly smaller number of components
are considered at each update of the cluster assignment variables, and
thus the algorithm is faster and has lower ESSs. If a single slice
variable is used instead, as proposed in Griffin and Walker~\cite
{GriWal2011a}, or if a nonslice conditional sampler like Barrios et al.~\cite{BarLijNie2012a} is used, then all instantiated components will be
considered at each update. This can result in not only higher ESSs but
also higher computational overheads since the number of empty
components introduced can be very large. The bottom half of Figure~\ref{fig:numclusters} shows the distribution of the number of empty
components for one of the runs for the nonconjugate case (on the
logarithmic scale). The mean numbers of empty components are $76.6$ and
$31.4$ for the galaxy and acidity data sets, respectively. Other runs
and the conjugate case are similar and not shown. For comparison, the
top panels of Figure~\ref{fig:numclusters}
show the posterior
distribution of the number of nonempty components, which are smaller.
As a further note, we have found that the truncation of the slice
variables at $10^{-8}$ described in Section~\ref{sec:slice} is
essential to the program working properly, as otherwise it will
sometimes generate far too many atoms, causing the program to run out
of memory. Table~\ref{table:truncation} shows the number of times the
truncation came into effect during each MCMC run. We did not find cases
in which the $10^6$ limit on the number of atoms was reached among
these runs.

%
\begin{table}[b]
\tabcolsep=0pt
\caption{Average number of times the slice threshold $S$ was less than
the $10^{-8}$ truncation level over the 10 conditional slice sampling
runs. The total number of iterations of each run is 210,000. Each entry
reports the average and standard error over 10 runs}\label{table:truncation}
\begin{tabular*}{\columnwidth}{@{\extracolsep{\fill}}lcccc@{}}
\hline
\textbf{Model} & {\textbf{Galaxy}} & \textbf{Acidity} & \textbf{Geyser} & \textbf{Spikes}\\
\hline
CM & $4476\pm440$ & $6143\pm1148$ &--&--\\
CS & $4597\pm385$ & $4385\pm394$ &--&--\\
NS & $3712\pm222$ & $8017\pm1180$
& $15\mbox{,}180\pm980$ & $5621\pm475$ \\
\hline
\end{tabular*}
\end{table}
%

%
%
\begin{table*}[b]
\tabcolsep=0pt
\caption{Comparison of sampler efficiencies on the geyser (2D) and
spikes (6D) data sets. Each of 10 runs produces 10,000 samples, at
intervals of 20 iterations, after an initial burn-in period of 10,000
iterations. Each entry reports the average and standard error over the
10 runs}\label{table:efficienciesmv}
\begin{tabular*}{\textwidth}{@{\extracolsep{\fill}}lccccc@{}}
\hline
& &\multicolumn{2}{c}{\textbf{Geyser}} & \multicolumn
{2}{c}{\textbf{Spikes}} \\
\ccline{3-4,5-6}
\textbf{Model}& \textbf{Sampler} & \textbf{Runtime (s)} & \textbf{ESS} & \textbf{Runtime (s)} & \textbf{ESS} \\
\hline
NS & Cond Slice&
$142.6\pm1.1$ & $ 574\pm36$ &$732.6\pm8.1$ & $17.1\pm2.3$ \\
NS & Marg Reuse ($C=1$) &
$208.0\pm1.3$ & $2770\pm209$ &$1120.3\pm8.8$ & $35.7\pm2.4$ \\
NS & Marg Reuse ($C=2$) &
$225.3\pm1.4$ & $3236\pm73$ &$1164.5\pm5.4$ & $46.9\pm2.9$ \\
NS & Marg Reuse ($C=3$) &
$241.5\pm1.3$ & $3148\pm71$ &$1204.1\pm7.3$ & $57.0\pm3.9$ \\
NS & Marg Reuse ($C=4$) &
$257.7\pm1.7$ & $3291\pm145$ &$1238.5\pm7.8$ & $61.4\pm3.3$ \\
NS & Marg Reuse ($C=5$) &
$274.8\pm1.7$ & $3144\pm70$ &$1291.8\pm7.9$ & $69.8\pm4.9$ \\
NS & Marg Reuse ($C=10$) &
$356.3\pm2.5$ & $3080\pm135$ &$1513.8\pm11.9$ & $90.8\pm5.6$ \\
NS & Marg Reuse ($C=15$) &
$446.6\pm4.9$ & $3312\pm154$ &$1746.3\pm10.7$ & $95.9\pm4.2$ \\
NS & Marg Reuse ($C=20$) &
$550.4\pm3.5$ & $3336\pm109$ &$1944.0\pm14.7$ & $114.5\pm8.4$ \\
\hline
\end{tabular*}
\end{table*}

\subsection{Multidimensional Data Sets: Geyser and Spikes}

We have also explored the efficacies of the algorithms on the geyser
and spikes data sets. For the spikes data set we reduced the size of
the data set by randomly selecting 500 spike waveforms to reduce the
overall computation time for the experiments. In preliminary
experiments this does not affect the qualitative conclusions drawn from
the results. We did not include the generalization of Neal's Algorithm
8 in these experiments, as we have found in initial explorations that
it took significantly more computation time without producing
substantially higher ESSs than the Reuse algorithm. The setups of the
experiments are similar as for the one-dimensional setting, with each
algorithm producing ten independent runs, each consisting of 10,000
burn-in iterations followed by 10,000 samples collected at intervals of
20 iterations. In addition to $C=1,\ldots,5$, we also explored higher
values of $C=10$, $15$ and $20$.

\begin{figure*}

\includegraphics{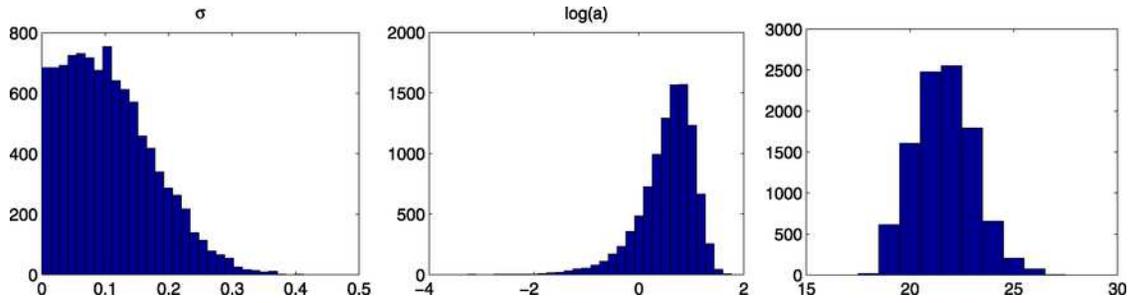}

\caption{Histograms of the posterior distribution of $\sigma$, $\log
(a)$ and the number of clusters, respectively, for the spikes
data.}\label{fig:spikesposterior}
\end{figure*}

The run times and ESSs are reported in Table~\ref{table:efficienciesmv}. The trends observed for the one-dimensional
setting hold here as well: that the slice sampler is faster but
produces lower ESSs, and that with increasing $C$ the marginalized
sampler produces higher ESSs at higher computational costs. As
expected, the algorithms mix more slowly on the higher-dimensional
spikes data set, with significantly lower ESSs. For the spikes data set
the nonconjugate marginalized samplers with higher values of $C$ have
significantly higher ESSs. In fact, they had better ESSs per unit of
run time than for lower values of $C$ or for the slice sampler. This
contrasts with the other simpler data sets, where lower values of $C$
worked very well, probably because the additional complexity of higher
$C$ values was not needed. Figure~\ref{fig:spikesposterior} shows the
posterior distributions over the number of clusters, $\log(a)$ and~$\sigma$.

%

%
\begin{figure*}

\includegraphics{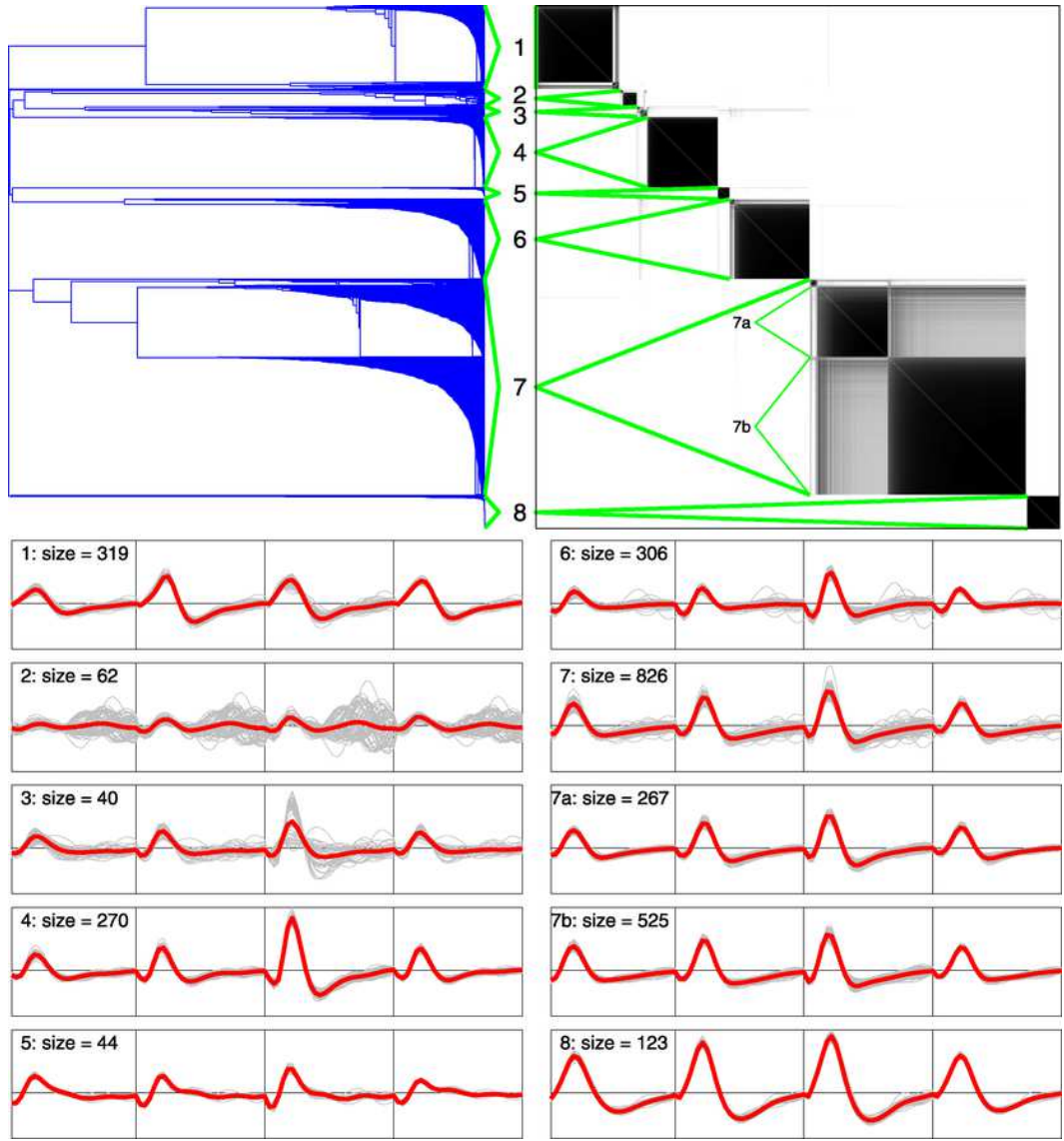}

\caption{Top: hierarchical organization of spike waveforms obtained by
average linkage and the corresponding reordered co-clustering matrix.
Bottom: clusters found by thresholding at $0.95$. Each panel consists
of four subpanels, each corresponding to the waveforms recorded by an
electrode. Each waveform in the cluster is plotted in light grey and their
mean in dark grey.}\label{fig:spikes}
\end{figure*}

Finally, we illustrate the clustering structure among spike waveforms
discovered by the NGGP mixture model. 2000 spike waveforms were
selected at random from the data set and the Reuse algorithm with
$C=20$ is run as before, with 10,000 burn-in iterations followed by
10,000 samples collected every 20 iterations. We use co-clustering
probabilities to summarize the clustering structure. For each pair
$(i,j)$ of spikes let $p_{ij}$ be the (estimated) posterior probability
that the two spikes were assigned to the same cluster. We use average
linkage to organize the spikes into a hierarchy, where the distance
between spikes $i$ and $j$ is defined to be $1-p_{ij}$. This is then
used to reorder the co-clustering matrix. The hierarchy and reordered
matrix are shown on the upper panels of Figure~\ref{fig:spikes}.
We see that most spikes belong to six large clusters, two of which have
significant overlap and merged into one, while a subset of waveforms
formed smaller clusters which may or may not overlap with other
clusters. In the bottom panels of Figure~\ref{fig:spikes} we visualize
the various clusters found by thresholding the hierarchy at $0.95$ and
ignoring clusters of size less than 10.

We can interpret the clusters found here in the context of spike
sorting, an important process in experimental neuroscience of detecting
spikes from neural recordings and determining the neuron\vadjust{\goodbreak} corresponding
to each spike from the shape of its waveform (as well as the number of
neurons) using a variety of manual or automated clustering techniques,
with each cluster interpreted as a unique neuron. See Quiroga~\cite
{Qui2007a} and Lewicki~\cite{Lew1998a} for reviews of spike sorting
methods, and also G\"or\"ur and Rasmussen~\cite{GorRas2004a}, Wood and
Black~\cite{WooBla2008a} and Gasthaus et al.~\cite{GasWooGor2009a} for
Bayesian nonparametric mixture modeling ap-\break proaches to spike sorting. We
find that the 5 largest clusters found (1, 4, 6, 7 and 8) all
correspond to well-defined waveforms with distinctive shapes, and
expect each of 1, 4, 6 and 8 to correspond to a single neuron. Spikes
in cluster 5 have similar waveforms, as 6 and the two clusters are in
fact merged at a threshold of 0.99, though spikes in 5 lacked refractory
periods; they may either correspond to the same or distinct neurons.
Clusters 2 and 3 consist of outliers, false detections or waveforms
formed by the superposition of two consecutive spikes. We note that a
number of waveforms in other clusters are also superpositions as well.
Finally, analyzing the two subclusters of~7, we see that although their
shapes are very similar, the waveforms in the first two subpanels of 7a
seem to be slightly smaller than those in 7b, though it is unclear if
the subclustering is due to two neurons or is an artefact of the
mixture components not being flexible enough to capture spike waveform
variability.

The approach taken here is simply to use an agglomerative linkage
algorithm to help us visualize and explore the posterior over partition
structures under the mixture model. An alternative approach is to
summarize the posterior using a single partition, for example, using
the maximum a posteriori partition or one that minimizes the posterior
expectation of a loss function like Binder's loss. The issue of how
best to analyze and interpret the posterior partition structure of
Bayesian models for clustering is still an open question and beyond the
scope of this paper. We refer the interested reader to Binder~\cite
{Bin1978a}, Medvedovic and Sivaganesan~\cite{MedSiv2002a}, Dahl~\cite
{Dah2006a}, Lau and Green~\cite{LauGre2007a}, Fritsch and Ickstadt
\cite
{FriIck2009a} and Rasmussen et al.~\cite{RasDe-Gha2009a} for classical
and recent efforts in this regard.


\section{Discussion}\label{Discussion_Section}

NRMs provide a large class of flexible nonparametric priors beyond the
standard DP, but their more common use is currently hindered by a lack
of understanding and of good algorithms for posterior simulation. This
work provides a review of NRMs for easy access\vadjust{\goodbreak} to the extensive
literature, as well as novel algorithms for posterior simulation that
are efficient and easy to use. We will also provide open source Java
software implementing all four algorithms described in Section~\ref{MCMC_section} so that others might more easily explore them.

All the samplers proposed in this paper are basic samplers that make
changes to the cluster assignment of one observation at a time.
Samplers that make more complex changes, for example, those based on
split-merge Metropolis--Hastings moves by Jain and Neal~\cite
{JaiNea2000a}, can be significantly more efficient at exploring
multiple posterior modes. Such samplers can be derived in both
marginalized and conditional forms, using the characterizations
reviewed in this paper, and are an interesting avenue of future
research. Beyond the algorithms described in Section~\ref{MCMC_section}, there are many variants possible with both marginalized
and conditional samplers for NRM mixture models. While conditional
samplers have been well explored in the literature, ours are the first
tractable marginalized samplers for mixture models with a homogeneous
NRM prior. In addition, a~number of samplers based on the system of
predictive distributions of NRMs have been proposed by James et al.~\cite{JamLijPru2009a} and by Lijoi et al.~\cite
{LijMenPru2005b,LijMenPru2005a,LijMenPru2007a}, but these sampling
methods can be computationally expensive in the nonconjugate setting
due to numerical integrations needed for computing the probabilities
associated to new clusters, and convergence is slow, requiring
additional acceleration steps. See, for example, Bush and MacEachern
\cite{BusMac1996a} for details.

A random probability measure that is in popular use but conspicuously
not within the class of NRMs is the two-parameter Poisson--Dirichlet
process (otherwise known as the Pitman--Yor process) by Perman et al.
\cite{PerPitYor1992a}. See also Pitman and Yor~\cite{PitYor1997a} and
Ishwaran and James~\cite{IshJam2001a} for details. It is instead in an
even larger class known as the Poisson--Kingman processes introduced by
Pitman~\cite{Pit2003a}, which are obtained by allowing the total mass
of the otherwise completely random measure underlying the NRM to have a
different distribution. Poisson--Kingman processes represent the largest
known class of random probability measures that are still
mathematically tractable. In addition to NRMs, they also include random
probability measures induced by the so-called Gibbs type exchangeable
random partitions introduced by Gnedin and Pitman~\cite{GnePit2006a}.
The marginalized and conditional samplers we have developed may be
extended to the Poisson--Kingman processes as well.

Throughout this paper we have used the NGGP as a running example to
illustrate the various properties and formulae, because of its
tractability and because it includes many well-known NRMs as examples.
It has been shown by Lijoi et al.~\cite{LijPruWal2008a} that the NGGP
is the only NRM that is also of Gibbs type. Beyond the NGGP, the
formulae derived tend to become intractable and require numerical
integrations. A notable exception is the class of NRMs whose L\'evy
intensity measure are mixtures of those for the generalized Gamma CRM,
first proposed by Trippa and Favaro~\cite{TriFav2011a} who also showed
that they form a dense subclass of the NRMs. It is straightforward to
extend the algorithms and the software derived in this paper to this
larger class.

As a final remark, the study of random probability measures underpins a
large body of work spanning probability, statistics, combinatorics and
mathematical genetics. They also form the core of many Bayesian
nonparametric models that are increasingly popular in applied
statistics and machine learning. By expanding the class of tractable
random probability measures beyond the DP to NRMs, we hope that our
work will increase both the range and flexibility of the models in use
now and in the future.


\section*{Acknowledgments}

The authors are grateful to the Editor, an Associate Editor and two
anonymous referees for their constructive comments and suggestions.
This work was supported by the European Research Council (ERC) through
StG ``N-BNP'' 306406.
%
%

%




\begin{thebibliography}{87}

\bibitem{Ald1985a}
\begin{bincollection}[mr]
\bauthor{\bsnm{Aldous},~\bfnm{David~J.}\binits{D.~J.}}
(\byear{1985}).
\btitle{Exchangeability and related topics}.
In \bbooktitle{\'{E}cole D'\'et\'e de Probabilit\'es de {S}aint-{F}lour,
  {XIII}---1983}.
\bseries{Lecture Notes in Math.}
\bvolume{1117}
\bpages{1--198}.
\bpublisher{Springer}, \blocation{Berlin}.
\bid{doi={10.1007/BFb0099421}, mr={0883646}}
\bptok{imsref}%
\end{bincollection}
\endbibitem

\bibitem{BarLijNie2012a}
\begin{bmisc}[auto:STB|2013/06/05|13:45:01]
\bauthor{\bsnm{Barrios},~\bfnm{E.}\binits{E.}},
  \bauthor{\bsnm{Lijoi},~\bfnm{A.}\binits{A.}},
  \bauthor{\bsnm{Nieto-Barajas},~\bfnm{L.~E.}\binits{L.~E.}} \AND
  \bauthor{\bsnm{Pr{\"u}enster},~\bfnm{I.}\binits{I.}}
(\byear{2012}).
\bhowpublished{Modeling with normalized random measure mixture models.
  Unpublished manuscript}.
\bptok{imsref}%
\end{bmisc}
\endbibitem

\bibitem{Bin1978a}
\begin{barticle}[mr]
\bauthor{\bsnm{Binder},~\bfnm{D.~A.}\binits{D.~A.}}
(\byear{1978}).
\btitle{Bayesian cluster analysis}.
\bjournal{Biometrika}
\bvolume{65}
\bpages{31--38}.
\bid{issn={0006-3444}, mr={0501592}}
\bptok{imsref}%
\end{barticle}
\endbibitem

\bibitem{BlaMac1973a}
\begin{barticle}[mr]
\bauthor{\bsnm{Blackwell},~\bfnm{David}\binits{D.}} \AND
  \bauthor{\bsnm{MacQueen},~\bfnm{James~B.}\binits{J.~B.}}
(\byear{1973}).
\btitle{Ferguson distributions via {P}\'olya urn schemes}.
\bjournal{Ann. Statist.}
\bvolume{1}
\bpages{353--355}.
\bid{issn={0090-5364}, mr={0362614}}
\bptok{imsref}%
\end{barticle}
\endbibitem

\bibitem{Bri1999a}
\begin{barticle}[mr]
\bauthor{\bsnm{Brix},~\bfnm{Anders}\binits{A.}}
(\byear{1999}).
\btitle{Generalized gamma measures and shot-noise {C}ox processes}.
\bjournal{Adv. in Appl. Probab.}
\bvolume{31}
\bpages{929--953}.
\bid{doi={10.1239/aap/1029955251}, issn={0001-8678}, mr={1747450}}
\bptok{imsref}%
\end{barticle}
\endbibitem

\bibitem{BroJorPit2012a}
\begin{bmisc}[mr]
\bauthor{\bsnm{Broderick},~\bfnm{Tamara}\binits{T.}},
  \bauthor{\bsnm{Jordan},~\bfnm{Michael~I.}\binits{M.~I.}} \AND
  \bauthor{\bsnm{Pitman},~\bfnm{Jim}\binits{J.}}
(\byear{2012}).
\bhowpublished{Clusters and features from combinatorial stochastic processes. Available at \arxivurl{arXiv:1206.5862} [math.ST].}
\bptok{imsref}%
\end{bmisc}
\endbibitem

\bibitem{BusMac1996a}
\begin{barticle}[auto:STB|2013/06/05|13:45:01]
\bauthor{\bsnm{Bush},~\bfnm{C.~A.}\binits{C.~A.}} \AND
  \bauthor{\bsnm{MacEachern},~\bfnm{S.~N.}\binits{S.~N.}}
(\byear{1996}).
\btitle{A semiparametric Bayesian model for randomised block designs}.
\bjournal{Biometrika}
\bvolume{83}
\bpages{275--285}.
\bptok{imsref}%
\end{barticle}
\endbibitem

\bibitem{Dah2006a}
\begin{bincollection}[auto:STB|2013/06/05|13:45:01]
\bauthor{\bsnm{Dahl},~\bfnm{D.~B.}\binits{D.~B.}}
(\byear{2006}).
\btitle{Model-based clustering for expression data via a Dirichlet process
  mixture model}.
In \bbooktitle{Bayesian Inference for Gene Expression and Proteomics}
(\beditor{\bfnm{K.}\binits{K.}~\bsnm{Do}},
  \beditor{\bfnm{P.}\binits{P.}~\bsnm{M{\"u}ller}} \AND
  \beditor{\bfnm{M.}\binits{M.}~\bsnm{Vannucci}}, eds.).
\bpublisher{Cambridge Univ. Press}, \blocation{Cambridge}.
\bptok{imsref}%
\end{bincollection}
\endbibitem

\bibitem{DalVer2008a}
\begin{bbook}[mr]
\bauthor{\bsnm{Daley},~\bfnm{D.~J.}\binits{D.~J.}} \AND
  \bauthor{\bsnm{Vere-Jones},~\bfnm{D.}\binits{D.}}
(\byear{2002}).
\btitle{An Introduction to the Theory of Point Processes}.
\bpublisher{Springer}, \blocation{New York}.
\bid{mr={0950166}}
\bptnote{check year}%
\bptok{imsref}%
\end{bbook}
\endbibitem

\bibitem{DieRob94}
\begin{barticle}[mr]
\bauthor{\bsnm{Diebolt},~\bfnm{Jean}\binits{J.}} \AND
  \bauthor{\bsnm{Robert},~\bfnm{Christian~P.}\binits{C.~P.}}
(\byear{1994}).
\btitle{Estimation of finite mixture distributions through {B}ayesian
  sampling}.
\bjournal{J. R. Stat. Soc. Ser. B Stat. Methodol.}
\bvolume{56}
\bpages{363--375}.
\bid{issn={0035-9246}, mr={1281940}}
\bptok{imsref}%
\end{barticle}
\endbibitem

\bibitem{Esc88}
\begin{bmisc}[mr]
\bauthor{\bsnm{Escobar},~\bfnm{Michael~David}\binits{M.~D.}}
(\byear{1988}).
\bhowpublished{Estimating the means of several normal populations by
  nonparametric estimation of the distribution of the means. Ph.D. thesis, Yale Univ}.
\bid{mr={2637324}}
\bptok{imsref}%
\end{bmisc}
\endbibitem

\bibitem{Esc94}
\begin{barticle}[mr]
\bauthor{\bsnm{Escobar},~\bfnm{Michael~D.}\binits{M.~D.}}
(\byear{1994}).
\btitle{Estimating normal means with a {D}irichlet process prior}.
\bjournal{J. Amer. Statist. Assoc.}
\bvolume{89}
\bpages{268--277}.
\bid{issn={0162-1459}, mr={1266299}}
\bptok{imsref}%
\end{barticle}
\endbibitem

\bibitem{EscWes1995a}
\begin{barticle}[mr]
\bauthor{\bsnm{Escobar},~\bfnm{Michael~D.}\binits{M.~D.}} \AND
  \bauthor{\bsnm{West},~\bfnm{Mike}\binits{M.}}
(\byear{1995}).
\btitle{Bayesian density estimation and inference using mixtures}.
\bjournal{J.~Amer. Statist. Assoc.}
\bvolume{90}
\bpages{577--588}.
\bid{issn={0162-1459}, mr={1340510}}
\bptok{imsref}%
\end{barticle}
\endbibitem

\bibitem{Ewe1972a}
\begin{barticle}[mr]
\bauthor{\bsnm{Ewens},~\bfnm{W.~J.}\binits{W.~J.}}
(\byear{1972}).
\btitle{The sampling theory of selectively neutral alleles}.
\bjournal{Theoret. Population Biology}
\bvolume{3}
\bpages{87--112; erratum, ibid. \textbf{3} (1972), 240, 376}.
\bid{issn={0040-5809}, mr={0325177}}
\bptnote{check related}%
\bptok{imsref}%
\end{barticle}
\endbibitem

\bibitem{Fav12}
\begin{bmisc}[auto:STB|2013/06/05|13:45:01]
\bauthor{\bsnm{Favaro},~\bfnm{S.}\binits{S.}} \AND
  \bauthor{\bsnm{Walker},~\bfnm{S.~G.}\binits{S.~G.}}
(\byear{2013}).
\bhowpublished{Slice sampling {$\sigma $}-stable Poisson--{K}ingman mixture
  models. \textit{J. Comput. Graph. Statist.} To appear}.
\bptok{imsref}%
\end{bmisc}
\endbibitem

\bibitem{Fer1973a}
\begin{barticle}[mr]
\bauthor{\bsnm{Ferguson},~\bfnm{Thomas~S.}\binits{T.~S.}}
(\byear{1973}).
\btitle{A {B}ayesian analysis of some nonparametric problems}.
\bjournal{Ann. Statist.}
\bvolume{1}
\bpages{209--230}.
\bid{issn={0090-5364}, mr={0350949}}
\bptok{imsref}%
\end{barticle}
\endbibitem

\bibitem{FerKla1972a}
\begin{barticle}[mr]
\bauthor{\bsnm{Ferguson},~\bfnm{Thomas~S.}\binits{T.~S.}} \AND
  \bauthor{\bsnm{Klass},~\bfnm{Michael~J.}\binits{M.~J.}}
(\byear{1972}).
\btitle{A representation of independent increment processes without {G}aussian
  components}.
\bjournal{Ann. Math. Statist.}
\bvolume{43}
\bpages{1634--1643}.
\bid{issn={0003-4851}, mr={0373022}}
\bptok{imsref}%
\end{barticle}
\endbibitem

\bibitem{FraRaf2007a}
\begin{barticle}[mr]
\bauthor{\bsnm{Fraley},~\bfnm{Chris}\binits{C.}} \AND
  \bauthor{\bsnm{Raftery},~\bfnm{Adrian~E.}\binits{A.~E.}}
(\byear{2007}).
\btitle{Bayesian regularization for normal mixture estimation and model-based
  clustering}.
\bjournal{J.~Classification}
\bvolume{24}
\bpages{155--181}.
\bid{doi={10.1007/s00357-007-0004-5}, issn={0176-4268}, mr={2415725}}
\bptok{imsref}%
\end{barticle}
\endbibitem

\bibitem{FriIck2009a}
\begin{barticle}[mr]
\bauthor{\bsnm{Fritsch},~\bfnm{Arno}\binits{A.}} \AND
  \bauthor{\bsnm{Ickstadt},~\bfnm{Katja}\binits{K.}}
(\byear{2009}).
\btitle{Improved criteria for clustering based on the posterior similarity
  matrix}.
\bjournal{Bayesian Anal.}
\bvolume{4}
\bpages{367--391}.
\bid{doi={10.1214/09-BA414}, issn={1936-0975}, mr={2507368}}
\bptok{imsref}%
\end{barticle}
\endbibitem

\bibitem{GasWooGor2009a}
\begin{bincollection}[auto:STB|2013/06/05|13:45:01]
\bauthor{\bsnm{Gasthaus},~\bfnm{J.}\binits{J.}},
  \bauthor{\bsnm{Wood},~\bfnm{F.}\binits{F.}},
  \bauthor{\bsnm{G{\"o}r{\"u}r},~\bfnm{D.}\binits{D.}} \AND
  \bauthor{\bsnm{Teh},~\bfnm{Y.~W.}\binits{Y.~W.}}
(\byear{2009}).
\btitle{Dependent Dirichlet process spike sorting}.
In \bbooktitle{Advances in Neural Information Processing Systems \textit{21}}
\bpages{497--504}.
\bptok{imsref}%
\end{bincollection}
\endbibitem

\bibitem{GilWil1992a}
\begin{barticle}[auto:STB|2013/06/05|13:45:01]
\bauthor{\bsnm{Gilks},~\bfnm{W.~R.}\binits{W.~R.}} \AND
  \bauthor{\bsnm{Wild},~\bfnm{P.}\binits{P.}}
(\byear{1992}).
\btitle{Adaptive rejection sampling for Gibbs sampling}.
\bjournal{Appl. Statist.}
\bvolume{41}
\bpages{337--348}.
\bptok{imsref}%
\end{barticle}
\endbibitem

\bibitem{GnePit2006a}
\begin{barticle}[mr]
\bauthor{\bsnm{Gnedin},~\bfnm{A.}\binits{A.}} \AND
  \bauthor{\bsnm{Pitman},~\bfnm{J.}\binits{J.}}
(\byear{2006}).
\btitle{Exchangeable {G}ibbs partitions and {S}tirling triangles}.
\bjournal{J. Math. Sci.}
\bvolume{138}
\bpages{5674--5684}.
\bptok{imsref}%
\end{barticle}
\endbibitem

\bibitem{GorRas2004a}
\begin{bmisc}[auto:STB|2013/06/05|13:45:01]
\bauthor{\bsnm{G{\"o}r{\"u}r},~\bfnm{D.}\binits{D.}},
\bauthor{\bsnm{Rasmussen},~\bfnm{C.~E.}\binits{C.~E.}},
\bauthor{\bsnm{Tolias},~\bfnm{A.~S.}\binits{A.~S.}},
\bauthor{\bsnm{Sinz},~\bfnm{F.}\binits{F.}}
\AND
\bauthor{\bsnm{Logothetis},~\bfnm{N.~K.}\binits{N.~K.}}
(\byear{2004}).
\bhowpublished{Modelling spikes with mixtures of factor analysers. In
  \textit{Proceedings of the Conference of the German Association for Pattern
  Recognition (DAGM)}.}
\bptok{imsref}%
\end{bmisc}
\endbibitem


\bibitem{Gre1995a}
\begin{barticle}[mr]
\bauthor{\bsnm{Green},~\bfnm{Peter~J.}\binits{P.~J.}}
(\byear{1995}).
\btitle{Reversible jump {M}arkov chain {M}onte {C}arlo computation and
  {B}ayesian model determination}.
\bjournal{Biometrika}
\bvolume{82}
\bpages{711--732}.
\bid{doi={10.1093/biomet/82.4.711}, issn={0006-3444}, mr={1380810}}
\bptok{imsref}%
\end{barticle}
\endbibitem

\bibitem{Gre01}
\begin{barticle}[mr]
\bauthor{\bsnm{Green},~\bfnm{Peter~J.}\binits{P.~J.}} \AND
  \bauthor{\bsnm{Richardson},~\bfnm{Sylvia}\binits{S.}}
(\byear{2001}).
\btitle{Modelling heterogeneity with and without the {D}irichlet process}.
\bjournal{Scand. J. Stat.}
\bvolume{28}
\bpages{355--375}.
\bid{doi={10.1111/1467-9469.00242}, issn={0303-6898}, mr={1842255}}
\bptok{imsref}%
\end{barticle}
\endbibitem

\bibitem{GriKolSte2011a}
\begin{barticle}[auto:STB|2013/06/05|13:45:01]
\bauthor{\bsnm{Griffin},~\bfnm{J.~E.}\binits{J.~E.}},
  \bauthor{\bsnm{Kolossiatis},~\bfnm{M.}\binits{M.}} \AND
  \bauthor{\bsnm{Steel},~\bfnm{M.~F.~J.}\binits{M.~F.~J.}}
(\byear{2013}).
\btitle{Comparing distributions using dependent normalized random
  measure mixtures}.
\bjournal{J. R. Stat. Soc. Ser. B Stat. Methodol.}
\bvolume{75}
\bpages{499--529}.
\bptok{imsref}%
\end{barticle}
\endbibitem

\bibitem{GriWal2011a}
\begin{barticle}[mr]
\bauthor{\bsnm{Griffin},~\bfnm{Jim~E.}\binits{J.~E.}} \AND
  \bauthor{\bsnm{Walker},~\bfnm{Stephen~G.}\binits{S.~G.}}
(\byear{2011}).
\btitle{Posterior simulation of normalized random measure mixtures}.
\bjournal{J. Comput. Graph. Statist.}
\bvolume{20}
\bpages{241--259}.\break
\bid{doi={10.1198/jcgs.2010.08176}, issn={1061-8600}, mr={2816547}}
\bptok{imsref}%
\end{barticle}
\endbibitem

\bibitem{GriGha2011a}
\begin{barticle}[mr]
\bauthor{\bsnm{Griffiths},~\bfnm{Thomas~L.}\binits{T.~L.}} \AND
  \bauthor{\bsnm{Ghahramani},~\bfnm{Zoubin}\binits{Z.}}
(\byear{2011}).
\btitle{The {I}ndian buffet process: An introduction and review}.
\bjournal{J.~Mach. Learn. Res.}
\bvolume{12}
\bpages{1185--1224}.
\bid{issn={1532-4435}, mr={2804598}}
\bptok{imsref}%
\end{barticle}
\endbibitem

\bibitem{Hjo1990a}
\begin{barticle}[mr]
\bauthor{\bsnm{Hjort},~\bfnm{Nils~Lid}\binits{N.~L.}}
(\byear{1990}).
\btitle{Nonparametric {B}ayes estimators based on beta processes in models for
  life history data}.
\bjournal{Ann. Statist.}
\bvolume{18}
\bpages{1259--1294}.
\bid{doi={10.1214/aos/1176347749}, issn={0090-5364}, mr={1062708}}
\bptok{imsref}%
\end{barticle}
\endbibitem

\bibitem{HjoHolMul2010a}
\begin{bbook}[mr]
\beditor{\bsnm{Hjort},~\bfnm{Nils~Lid}\binits{N.~L.}},
  \beditor{\bsnm{Holmes},~\bfnm{Chris}\binits{C.}},
  \beditor{\bsnm{M{\"u}ller},~\bfnm{Peter}\binits{P.}} \AND
  \beditor{\bsnm{Wal\-ker},~\bfnm{Stephen~G.}\binits{S.~G.}}, eds.
(\byear{2010}).
\btitle{Bayesian Nonparametrics}.
\bseries{Cambridge Series in Statistical and Probabilistic Mathematics}
\bvolume{28}.
\bpublisher{Cambridge Univ. Press}, \blocation{Cambridge}.
\bid{doi={10.1017/CBO9780511802478}, mr={2722987}}
\bptok{imsref}%
\end{bbook}
\endbibitem

\bibitem{IshJam2001a}
\begin{barticle}[mr]
\bauthor{\bsnm{Ishwaran},~\bfnm{Hemant}\binits{H.}} \AND
  \bauthor{\bsnm{James},~\bfnm{Lancelot~F.}\binits{L.~F.}}
(\byear{2001}).
\btitle{Gibbs sampling methods for stick-breaking priors}.
\bjournal{J. Amer. Statist. Assoc.}
\bvolume{96}
\bpages{161--173}.
\bid{doi={10.1198/016214501750332758}, issn={0162-1459}, mr={1952729}}
\bptok{imsref}%
\end{barticle}
\endbibitem

\bibitem{JaiNea2000a}
\begin{bmisc}[auto:STB|2013/06/05|13:45:01]
\bauthor{\bsnm{Jain},~\bfnm{S.}\binits{S.}} \AND
  \bauthor{\bsnm{Neal},~\bfnm{R.~M.}\binits{R.~M.}}
(\byear{2000}).
\bhowpublished{A split-merge Markov chain Monte Carlo procedure for the
  Dirichlet process mixture model. Unpublished manuscript}.
\bptok{imsref}%
\end{bmisc}
\endbibitem

\bibitem{james}
\begin{bmisc}[auto]
\bauthor{\bsnm{James},~\bfnm{Lancelot~F.}\binits{L.~F.}}
(\byear{2002}).
\bhowpublished{Poisson process partition calculus with applications to exchangeable
models and Bayesian nonparametrics. Available at \arxivurl{arXiv:math/0205093v1}.}
\bptok{imsref}%
\end{bmisc}
\endbibitem


\bibitem{Lan2003a}
\begin{barticle}[mr]
\bauthor{\bsnm{James},~\bfnm{Lancelot~F.}\binits{L.~F.}}
(\byear{2003}).
\btitle{A simple proof of the almost sure discreteness of a class of random
  measures}.
\bjournal{Statist. Probab. Lett.}
\bvolume{65}
\bpages{363--368}.
\bid{doi={10.1016/j.spl.2003.08.005}, issn={0167-7152}, mr={2039881}}
\bptok{imsref}%
\end{barticle}
\endbibitem



\bibitem{JamLijPru2006a}
\begin{barticle}[mr]
\bauthor{\bsnm{James},~\bfnm{Lancelot~F.}\binits{L.~F.}},
  \bauthor{\bsnm{Lijoi},~\bfnm{Antonio}\binits{A.}} \AND
  \bauthor{\bsnm{Pr{\"u}nster},~\bfnm{Igor}\binits{I.}}
(\byear{2006}).
\btitle{Conjugacy as a distinctive feature of the {D}irichlet process}.
\bjournal{Scand.~J. Stat.}
\bvolume{33}
\bpages{105--120}.
\bid{doi={10.1111/j.1467-9469.2005.00486.x}, issn={0303-6898}, mr={2255112}}
\bptok{imsref}%
\end{barticle}
\endbibitem

\bibitem{JamLijPru2009a}
\begin{barticle}[mr]
\bauthor{\bsnm{James},~\bfnm{Lancelot~F.}\binits{L.~F.}},
  \bauthor{\bsnm{Lijoi},~\bfnm{Antonio}\binits{A.}} \AND
  \bauthor{\bsnm{Pr{\"u}nster},~\bfnm{Igor}\binits{I.}}
(\byear{2009}).
\btitle{Posterior analysis for normalized random measures with independent
  increments}.
\bjournal{Scand. J. Stat.}
\bvolume{36}
\bpages{76--97}.
\bid{doi={10.1111/j.1467-9469.2008.00609.x}, issn={0303-6898}, mr={2508332}}
\bptok{imsref}%
\end{barticle}
\endbibitem

\bibitem{JamLijPru2010a}
\begin{barticle}[mr]
\bauthor{\bsnm{James},~\bfnm{Lancelot~F.}\binits{L.~F.}},
  \bauthor{\bsnm{Lijoi},~\bfnm{Antonio}\binits{A.}} \AND
  \bauthor{\bsnm{Pr{\"u}nster},~\bfnm{Igor}\binits{I.}}
(\byear{2010}).
\btitle{On the posterior distribution of classes of random means}.
\bjournal{Bernoulli}
\bvolume{16}
\bpages{155--180}.
\bid{doi={10.3150/09-BEJ200}, issn={1350-7265}, mr={2648753}}
\bptok{imsref}%
\end{barticle}
\endbibitem

\bibitem{JasHolSte2005a}
\begin{barticle}[mr]
\bauthor{\bsnm{Jasra},~\bfnm{A.}\binits{A.}},
  \bauthor{\bsnm{Holmes},~\bfnm{C.~C.}\binits{C.~C.}} \AND
  \bauthor{\bsnm{Stephens},~\bfnm{D.~A.}\binits{D.~A.}}
(\byear{2005}).
\btitle{Markov chain {M}onte {C}arlo methods and the label switching problem in
  {B}ayesian mixture modeling}.
\bjournal{Statist. Sci.}
\bvolume{20}
\bpages{50--67}.
\bid{doi={10.1214/088342305000000016}, issn={0883-4237}, mr={2182987}}
\bptok{imsref}%
\end{barticle}
\endbibitem

\bibitem{KalGriWal2011a}
\begin{barticle}[mr]
\bauthor{\bsnm{Kalli},~\bfnm{Maria}\binits{M.}},
  \bauthor{\bsnm{Griffin},~\bfnm{Jim~E.}\binits{J.~E.}} \AND
  \bauthor{\bsnm{Walker},~\bfnm{Stephen~G.}\binits{S.~G.}}
(\byear{2011}).
\btitle{Slice sampling mixture models}.
\bjournal{Stat. Comput.}
\bvolume{21}
\bpages{93--105}.
\bid{doi={10.1007/s11222-009-9150-y}, issn={0960-3174}, mr={2746606}}
\bptok{imsref}%
\end{barticle}
\endbibitem

\bibitem{Kin1967a}
\begin{barticle}[mr]
\bauthor{\bsnm{Kingman},~\bfnm{J.~F.~C.}\binits{J.~F.~C.}}
(\byear{1967}).
\btitle{Completely random measures}.
\bjournal{Pacific J. Math.}
\bvolume{21}
\bpages{59--78}.
\bid{issn={0030-8730}, mr={0210185}}
\bptok{imsref}%
\end{barticle}
\endbibitem

\bibitem{Kin1993a}
\begin{bbook}[mr]
\bauthor{\bsnm{Kingman},~\bfnm{J.~F.~C.}\binits{J.~F.~C.}}
(\byear{1993}).
\btitle{Poisson Processes}.
\bseries{Oxford Studies in Probability}
\bvolume{3}.
\bpublisher{Clarendon Press}, \blocation{Oxford}.
\bid{mr={1207584}}
\bptok{imsref}%
\end{bbook}
\endbibitem

\bibitem{Kin1975a}
\begin{barticle}[mr]
\bauthor{\bsnm{Kingman},~\bfnm{J.~F.~C.}\binits{J.~F.~C.}},
  \bauthor{\bsnm{Taylor},~\bfnm{S.~J.}\binits{S.~J.}},
  \bauthor{\bsnm{Hawkes},~\bfnm{A.~G.}\binits{A.~G.}},
  \bauthor{\bsnm{Walker},~\bfnm{A.~M.}\binits{A.~M.}},
  \bauthor{\bsnm{Cox},~\bfnm{David~Roxbee}\binits{D.~R.}},
  \bauthor{\bsnm{Smith},~\bfnm{A.~F.~M.}\binits{A.~F.~M.}},
  \bauthor{\bsnm{Hill},~\bfnm{B.~M.}\binits{B.~M.}},
  \bauthor{\bsnm{Burville},~\bfnm{P.~J.}\binits{P.~J.}} \AND
  \bauthor{\bsnm{Leonard},~\bfnm{T.}\binits{T.}}
(\byear{1975}).
\btitle{Random discrete distribution}.
\bjournal{J. R. Stat. Soc. Ser. B Stat. Methodol.}
\bvolume{37}
\bpages{1--22}.
\bid{issn={0035-9246}, mr={0368264}}
\bptnote{check related}%
\bptok{imsref}%
\end{barticle}
\endbibitem

\bibitem{LauGre2007a}
\begin{barticle}[mr]
\bauthor{\bsnm{Lau},~\bfnm{John~W.}\binits{J.~W.}} \AND
  \bauthor{\bsnm{Green},~\bfnm{Peter~J.}\binits{P.~J.}}
(\byear{2007}).
\btitle{Bayesian model-based clustering procedures}.
\bjournal{J. Comput. Graph. Statist.}
\bvolume{16}
\bpages{526--558}.
\bid{doi={10.1198/106186007X238855}, issn={1061-8600}, mr={2351079}}
\bptok{imsref}%
\end{barticle}
\endbibitem

\bibitem{Lew1998a}
\begin{barticle}[auto:STB|2013/06/05|13:45:01]
\bauthor{\bsnm{Lewicki},~\bfnm{M.~S.}\binits{M.~S.}}
(\byear{1998}).
\btitle{A review of methods for spike sorting: The detection and classification
  of neural action potentials}.
\bjournal{Network}
\bvolume{9}
\bpages{53--78}.
\bptok{imsref}%
\end{barticle}
\endbibitem

\bibitem{LewShe1979a}
\begin{barticle}[mr]
\bauthor{\bsnm{Lewis},~\bfnm{P.~A.~W.}\binits{P.~A.~W.}} \AND
  \bauthor{\bsnm{Shedler},~\bfnm{G.~S.}\binits{G.~S.}}
(\byear{1979}).
\btitle{Simulation of nonhomogeneous {P}oisson processes by thinning}.
\bjournal{Naval Res. Logist. Quart.}
\bvolume{26}
\bpages{403--413}.
\bid{doi={10.1002/nav.3800260304}, issn={0028-1441}, mr={0546120}}
\bptok{imsref}%
\end{barticle}
\endbibitem

\bibitem{LijMenPru2005b}
\begin{barticle}[mr]
\bauthor{\bsnm{Lijoi},~\bfnm{Antonio}\binits{A.}},
  \bauthor{\bsnm{Mena},~\bfnm{Rams{\'e}s~H.}\binits{R.~H.}} \AND
  \bauthor{\bsnm{Pr{\"u}nster},~\bfnm{Igor}\binits{I.}}
(\byear{2005}).
\btitle{Bayesian nonparametric analysis for a generalized {D}irichlet process
  prior}.
\bjournal{Stat. Inference Stoch. Process.}
\bvolume{8}
\bpages{283--309}.
\bid{doi={10.1007/s11203-005-6071-z}, issn={1387-0874}, mr={2177315}}
\bptok{imsref}%
\end{barticle}
\endbibitem

\bibitem{LijMenPru2005a}
\begin{barticle}[mr]
\bauthor{\bsnm{Lijoi},~\bfnm{Antonio}\binits{A.}},
  \bauthor{\bsnm{Mena},~\bfnm{Rams{\'e}s~H.}\binits{R.~H.}} \AND
  \bauthor{\bsnm{Pr{\"u}nster},~\bfnm{Igor}\binits{I.}}
(\byear{2005}).
\btitle{Hierarchical mixture modeling with normalized inverse-{G}aussian
  priors}.
\bjournal{J.~Amer. Statist. Assoc.}
\bvolume{100}
\bpages{1278--1291}.
\bid{doi={10.1198/016214505000000132}, issn={0162-1459}, mr={2236441}}
\bptok{imsref}%
\end{barticle}
\endbibitem

\bibitem{LijMenPru2007a}
\begin{barticle}[mr]
\bauthor{\bsnm{Lijoi},~\bfnm{Antonio}\binits{A.}},
  \bauthor{\bsnm{Mena},~\bfnm{Rams{\'e}s~H.}\binits{R.~H.}} \AND
  \bauthor{\bsnm{Pr{\"u}nster},~\bfnm{Igor}\binits{I.}}
(\byear{2007}).
\btitle{Controlling the reinforcement in {B}ayesian non-parametric mixture
  models}.
\bjournal{J. R. Stat. Soc. Ser. B Stat. Methodol.}
\bvolume{69}
\bpages{715--740}.
\bid{doi={10.1111/j.1467-9868.2007.00609.x}, issn={1369-7412}, mr={2370077}}
\bptok{imsref}%
\end{barticle}
\endbibitem

\bibitem{LijPru2010a}
\begin{bincollection}[mr]
\bauthor{\bsnm{Lijoi},~\bfnm{Antonio}\binits{A.}} \AND
  \bauthor{\bsnm{Pr{\"u}nster},~\bfnm{Igor}\binits{I.}}
(\byear{2010}).
\btitle{Models beyond the {D}irichlet process}.
In \bbooktitle{Bayesian Nonparametrics}
(\beditor{\binits{N.~L.}\bfnm{N.~L.} \bsnm{Hjort}},
\beditor{\binits{C.~C.}\bfnm{C.~C.} \bsnm{Holmes}},
\beditor{\binits{P.}\bfnm{P.} \bsnm{M{\"u}ller}}
\AND
\beditor{\binits{S.~G.}\bfnm{S.~G.} \bsnm{Walker}}, eds.)
\bpages{80--136}.
\bpublisher{Cambridge Univ. Press}, \blocation{Cambridge}.
\bid{mr={2730661}}
\bptok{imsref}%
\end{bincollection}
\endbibitem

\bibitem{LijPruWal2008a}
\begin{barticle}[mr]
\bauthor{\bsnm{Lijoi},~\bfnm{Antonio}\binits{A.}},
  \bauthor{\bsnm{Pr{\"u}nster},~\bfnm{Igor}\binits{I.}} \AND
  \bauthor{\bsnm{Walker},~\bfnm{Stephen~G.}\binits{S.~G.}}
(\byear{2008}).
\btitle{Investigating nonparametric priors with {G}ibbs structure}.
\bjournal{Statist. Sinica}
\bvolume{18}
\bpages{1653--1668}.
\bid{issn={1017-0405}, mr={2469329}}
\bptok{imsref}%
\end{barticle}
\endbibitem

\bibitem{Lo1984a}
\begin{barticle}[mr]
\bauthor{\bsnm{Lo},~\bfnm{Albert~Y.}\binits{A.~Y.}}
(\byear{1984}).
\btitle{On a class of {B}ayesian nonparametric estimates. {I}. {D}ensity
  estimates}.
\bjournal{Ann. Statist.}
\bvolume{12}
\bpages{351--357}.
\bid{doi={10.1214/aos/1176346412}, issn={0090-5364}, mr={0733519}}
\bptok{imsref}%
\end{barticle}
\endbibitem

\bibitem{Mac1994a}
\begin{barticle}[mr]
\bauthor{\bsnm{MacEachern},~\bfnm{Steven~N.}\binits{S.~N.}}
(\byear{1994}).
\btitle{Estimating normal means with a conjugate style {D}irichlet process
  prior}.
\bjournal{Comm. Statist. Simulation Comput.}
\bvolume{23}
\bpages{727--741}.
\bid{doi={10.1080/03610919408813196}, issn={0361-0918}, mr={1293996}}
\bptok{imsref}%
\end{barticle}
\endbibitem


\bibitem{Mac1998a}
\begin{bincollection}[mr]
\bauthor{\bsnm{MacEachern},~\bfnm{Steven~N.}\binits{S.~N.}}
(\byear{1998}).
\btitle{Computational methods for mixture of {D}irichlet process models}.
In \bbooktitle{Practical Nonparametric and Semiparametric {B}ayesian
  Statistics}
(\beditor{\binits{D.}\bfnm{D.} \bsnm{Dey}},
\beditor{\binits{P.}\bfnm{P.} \bsnm{M{\"u}ller}}
\AND
\beditor{\binits{D.}\bfnm{D.} \bsnm{Sinha}}, eds.).
\bseries{Lecture Notes in Statist.}
\bvolume{133}
\bpages{23--43}.
\bpublisher{Springer}, \blocation{New York}.
\bid{doi={10.1007/978-1-4612-1732-9_2}, mr={1630074}}
\bptok{imsref}%
\end{bincollection}
\endbibitem

\bibitem{MacMul1998a}
\begin{barticle}[auto:STB|2013/06/05|13:45:01]
\bauthor{\bsnm{MacEachern},~\bfnm{S.~N.}\binits{S.~N.}} \AND
  \bauthor{\bsnm{M{\"u}ller},~\bfnm{P.}\binits{P.}}
(\byear{1998}).
\btitle{Estimating mixture of Dirichlet process models}.
\bjournal{J. Comput. Graph. Statist.}
\bvolume{7}
\bpages{223--238}.
\bptok{imsref}%
\end{barticle}
\endbibitem

\bibitem{mclbas88}
\begin{bbook}[mr]
\bauthor{\bsnm{McLachlan},~\bfnm{Geoffrey~J.}\binits{G.~J.}} \AND
  \bauthor{\bsnm{Basford},~\bfnm{Kaye~E.}\binits{K.~E.}}
(\byear{1988}).
\btitle{Mixture Models: Inference and Applications to Clustering}.
\bseries{Statistics: Textbooks and Monographs}
\bvolume{84}.
\bpublisher{Dekker}, \blocation{New York}.
\bid{mr={0926484}}
\bptok{imsref}%
\end{bbook}
\endbibitem

\bibitem{MedSiv2002a}
\begin{barticle}[auto:STB|2013/06/05|13:45:01]
\bauthor{\bsnm{Medvedovic},~\bfnm{M.}\binits{M.}} \AND
  \bauthor{\bsnm{Sivaganesan},~\bfnm{S.}\binits{S.}}
(\byear{2002}).
\btitle{Bayesian infinite mixture model based clustering of gene expression
  profiles}.
\bjournal{Bioinformatics}
\bvolume{18}
\bpages{1194--1206}.
\bptok{imsref}%
\end{barticle}
\endbibitem

\bibitem{MarRob96}
\begin{bincollection}[mr]
\bauthor{\bsnm{Mengersen},~\bfnm{K.~L.}\binits{K.~L.}} \AND
  \bauthor{\bsnm{Robert},~\bfnm{C.~P.}\binits{C.~P.}}
(\byear{1996}).
\btitle{Testing for mixtures: A {B}ayesian entropic approach}.
In \bbooktitle{Bayesian Statistics, 5 ({A}licante, 1994)}
(\beditor{\binits{J.~O.}\bfnm{J.~O.}~\bsnm{Berger}},
\beditor{\binits{J.~M.}\bfnm{J.~M.}~\bsnm{Bernardo}},
\beditor{\binits{A.~P.}\bfnm{A.~P.}~\bsnm{Dawid}},
\beditor{\binits{D.~V.}\bfnm{D.~V.}~\bsnm{Lindley}}
\AND
\beditor{\binits{A.~F.~M.}\bfnm{A.~F.~M.}~\bsnm{Smith}}, eds.)
\bpages{255--276}.
\bpublisher{Oxford Univ. Press}, \blocation{New York}.
\bid{mr={1425410}}
\bptok{imsref}%
\end{bincollection}
\endbibitem

\bibitem{multar98}
\begin{barticle}[mr]
\bauthor{\bsnm{Muliere},~\bfnm{Pietro}\binits{P.}} \AND
  \bauthor{\bsnm{Tardella},~\bfnm{Luca}\binits{L.}}
(\byear{1998}).
\btitle{Approximating distributions of random functionals of
  {F}erguson--{D}irichlet priors}.
\bjournal{Canad. J. Statist.}
\bvolume{26}
\bpages{283--297}.
\bid{doi={10.2307/3315511}, issn={0319-5724}, mr={1648431}}
\bptok{imsref}%
\end{barticle}
\endbibitem

\bibitem{MulErkWes1996a}
\begin{barticle}[mr]
\bauthor{\bsnm{M{\"u}ller},~\bfnm{Peter}\binits{P.}},
  \bauthor{\bsnm{Erkanli},~\bfnm{Alaattin}\binits{A.}} \AND
  \bauthor{\bsnm{West},~\bfnm{Mike}\binits{M.}}
(\byear{1996}).
\btitle{Bayesian curve fitting using multivariate normal mixtures}.
\bjournal{Biometrika}
\bvolume{83}
\bpages{67--79}.
\bid{doi={10.1093/biomet/83.1.67}, issn={0006-3444}, mr={1399156}}
\bptok{imsref}%
\end{barticle}
\endbibitem

\bibitem{Nea1992b}
\begin{bincollection}[auto:STB|2013/06/05|13:45:01]
\bauthor{\bsnm{Neal},~\bfnm{R.~M.}\binits{R.~M.}}
(\byear{1992}).
\btitle{Bayesian mixture modeling}.
In \bbooktitle{Proceedings of the 11th International Workshop on Maximum Entropy and Bayesian
Methods of Statistical Analysis, Seattle}.
\bpublisher{Kluwer}, \blocation{Dordrecht}.
\bptok{imsref}%
\end{bincollection}
\endbibitem

\bibitem{Nea2000a}
\begin{barticle}[mr]
\bauthor{\bsnm{Neal},~\bfnm{Radford~M.}\binits{R.~M.}}
(\byear{2000}).
\btitle{Markov chain sampling methods for {D}irichlet process mixture models}.
\bjournal{J. Comput. Graph. Statist.}
\bvolume{9}
\bpages{249--265}.
\bid{doi={10.2307/1390653}, issn={1061-8600}, mr={1823804}}
\bptok{imsref}%
\end{barticle}
\endbibitem

\bibitem{Nea2003a}
\begin{barticle}[mr]
\bauthor{\bsnm{Neal},~\bfnm{Radford~M.}\binits{R.~M.}}
(\byear{2003}).
\btitle{Slice sampling}.
\bjournal{Ann. Statist.}
\bvolume{31}
\bpages{705--767}.
\bid{doi={10.1214/aos/1056562461}, issn={0090-5364}, mr={1994729}}
\bptnote{check related}%
\bptok{imsref}%
\end{barticle}
\endbibitem

\bibitem{NiePru2009a}
\begin{barticle}[mr]
\bauthor{\bsnm{Nieto-Barajas},~\bfnm{Luis~E.}\binits{L.~E.}} \AND
  \bauthor{\bsnm{Pr{\"u}nster},~\bfnm{Igor}\binits{I.}}
(\byear{2009}).
\btitle{A sensitivity analysis for {B}ayesian nonparametric density
  estimators}.
\bjournal{Statist. Sinica}
\bvolume{19}
\bpages{685--705}.
\bid{issn={1017-0405}, mr={2514182}}
\bptok{imsref}%
\end{barticle}
\endbibitem

\bibitem{NiePruWal2004a}
\begin{barticle}[mr]
\bauthor{\bsnm{Nieto-Barajas},~\bfnm{Luis~E.}\binits{L.~E.}},
  \bauthor{\bsnm{Pr{\"u}nster},~\bfnm{Igor}\binits{I.}} \AND
  \bauthor{\bsnm{Wal\-ker},~\bfnm{Stephen~G.}\binits{S.~G.}}
(\byear{2004}).
\btitle{Normalized random measures driven by increasing additive processes}.
\bjournal{Ann. Statist.}
\bvolume{32}
\bpages{2343--2360}.
\bid{doi={10.1214/009053604000000625}, issn={0090-5364}, mr={2153987}}
\bptok{imsref}%
\end{barticle}
\endbibitem

\bibitem{Nob1994a}
\begin{bmisc}[mr]
\bauthor{\bsnm{Nobile},~\bfnm{Agostino}\binits{A.}}
(\byear{1994}).
\bhowpublished{Bayesian analysis of finite mixture distributions. Ph.D. thesis, Carnegie Mellon Univ}.
\bid{mr={2692049}}
\bptok{imsref}%
\end{bmisc}
\endbibitem

\bibitem{Oga1981a}
\begin{barticle}[auto:STB|2013/06/05|13:45:01]
\bauthor{\bsnm{Ogata},~\bfnm{Y.}\binits{Y.}}
(\byear{1981}).
\btitle{On Lewis' simulation method for Point processes}.
\bjournal{IEEE Trans. Inform. Theory}
\bvolume{27}
\bpages{23--31}.
\bptok{imsref}%
\end{barticle}
\endbibitem

\bibitem{Pap2008a}
\begin{bmisc}[auto:STB|2013/06/05|13:45:01]
\bauthor{\bsnm{Papaspiliopoulos},~\bfnm{O.}\binits{O.}}
 (\byear{2008}).
\bhowpublished{A note on posterior sampling from Dirichlet mixture models.
Working Paper 20, Centre for Research in Statistical Methodology, Univ. Warwick}.
\bptok{imsref}%
\end{bmisc}
\endbibitem

\bibitem{PapRob2008a}
\begin{barticle}[mr]
\bauthor{\bsnm{Papaspiliopoulos},~\bfnm{Omiros}\binits{O.}} \AND
  \bauthor{\bsnm{Roberts},~\bfnm{Gareth~O.}\binits{G.~O.}}
(\byear{2008}).
\btitle{Retrospective {M}arkov chain {M}onte {C}arlo methods for {D}irichlet
  process hierarchical models}.
\bjournal{Biometrika}
\bvolume{95}
\bpages{169--186}.
\bid{doi={10.1093/biomet/asm086}, issn={0006-3444}, mr={2409721}}
\bptok{imsref}%
\end{barticle}
\endbibitem

\bibitem{PerPitYor1992a}
\begin{barticle}[mr]
\bauthor{\bsnm{Perman},~\bfnm{Mihael}\binits{M.}},
  \bauthor{\bsnm{Pitman},~\bfnm{Jim}\binits{J.}} \AND
  \bauthor{\bsnm{Yor},~\bfnm{Marc}\binits{M.}}
(\byear{1992}).
\btitle{Size-biased sampling of {P}oisson point processes and excursions}.
\bjournal{Probab. Theory Related Fields}
\bvolume{92}
\bpages{21--39}.
\bid{doi={10.1007/BF01205234}, issn={0178-8051}, mr={1156448}}
\bptok{imsref}%
\end{barticle}
\endbibitem

\bibitem{Pit2003a}
\begin{bincollection}[mr]
\bauthor{\bsnm{Pitman},~\bfnm{Jim}\binits{J.}}
(\byear{2003}).
\btitle{Poisson--{K}ingman partitions}.
In \bbooktitle{Statistics and Science: A {F}estschrift for {T}erry {S}peed}
(\beditor{\binits{D.R.}\bfnm{D.~R.} \bsnm{Goldstein}}, ed.).
\bseries{Institute of Mathematical Statistics Lecture Notes---Monograph Series}
\bvolume{40}
\bpages{1--34}.
\bpublisher{IMS}, \blocation{Beachwood, OH}.
\bid{doi={10.1214/lnms/1215091133}, mr={2004330}}
\bptok{imsref}%
\end{bincollection}
\endbibitem

\bibitem{Pit2006a}
\begin{bbook}[mr]
\bauthor{\bsnm{Pitman},~\bfnm{J.}\binits{J.}}
(\byear{2006}).
\btitle{Combinatorial Stochastic Processes}.
\bseries{Lecture Notes in Math.}
\bvolume{1875}.
\bpublisher{Springer}, \blocation{Berlin}.
\bid{mr={2245368}}
\bptok{imsref}%
\end{bbook}
\endbibitem

\bibitem{PitYor1997a}
\begin{barticle}[mr]
\bauthor{\bsnm{Pitman},~\bfnm{Jim}\binits{J.}} \AND
  \bauthor{\bsnm{Yor},~\bfnm{Marc}\binits{M.}}
(\byear{1997}).
\btitle{The two-parameter {P}oisson--{D}irichlet distribution derived from a
  stable subordinator}.
\bjournal{Ann. Probab.}
\bvolume{25}
\bpages{855--900}.
\bid{doi={10.1214/aop/1024404422}, issn={0091-1798}, mr={1434129}}
\bptok{imsref}%
\end{barticle}
\endbibitem

\bibitem{Qui2007a}
\begin{barticle}[auto:STB|2013/06/05|13:45:01]
\bauthor{\bsnm{Quiroga},~\bfnm{R.~Q.}\binits{R.~Q.}}
(\byear{2007}).
\btitle{Spike sorting}.
\bjournal{Scholarpedia}
\bvolume{2}
\bpages{3583}.
\bptok{imsref}%
\end{barticle}
\endbibitem

\bibitem{Raf96}
\begin{bincollection}[auto:STB|2013/06/05|13:45:01]
\bauthor{\bsnm{Raftery},~\bfnm{A.~E.}\binits{A.~E.}}
(\byear{1996}).
\btitle{Hypothesis testing and model selection}.
In \bbooktitle{Markov Chain Monte Carlo in Practice}
(\beditor{\bfnm{W.~R.}\binits{W.~R.}~\bsnm{Gilks}},
  \beditor{\bfnm{S.}\binits{S.}~\bsnm{Richardson}} \AND
  \beditor{\bfnm{D.~J.}\binits{D.~J.}~\bsnm{Spiegelhalter}}, eds.).
\bpublisher{Chapman \& Hall}, \blocation{London}.
\bptok{imsref}%
\end{bincollection}
\endbibitem

\bibitem{Raf1996a}
\begin{bincollection}[auto:STB|2013/06/05|13:45:01]
\bauthor{\bsnm{Raftery},~\bfnm{A.~E.}\binits{A.~E.}}
(\byear{1996}).
\btitle{Hypothesis testing and model selection via posterior simulation}.
In \bbooktitle{Markov Chain Monte Carlo in Practice}
(\beditor{\bfnm{W.~R.}\binits{W.~R.}~\bsnm{Gilks}},
  \beditor{\bfnm{S.}\binits{S.}~\bsnm{Richardson}} \AND
  \beditor{\bfnm{D.~J.}\binits{D.~J.}~\bsnm{Spiegelhalter}}, eds.).
\bpublisher{Chapman \& Hall}, \blocation{London}.
\bptok{imsref}%
\end{bincollection}
\endbibitem

\bibitem{RasDe-Gha2009a}
\begin{barticle}[auto:STB|2013/06/05|13:45:01]
\bauthor{\bsnm{Rasmussen},~\bfnm{C.~E.}\binits{C.~E.}},
  \bauthor{\bparticle{De~la} \bsnm{Cruz},~\bfnm{B.~J.}\binits{B.~J.}},
  \bauthor{\bsnm{Ghahramani},~\bfnm{Z.}\binits{Z.}} \AND
  \bauthor{\bsnm{Wild},~\bfnm{D.~L.}\binits{D.~L.}}
(\byear{2009}).
\btitle{Modeling and visualizing uncertainty in gene expression clusters using
  Dirichlet process mixtures}.
\bjournal{IEEE/ACM Trans. Comput. Biol. and Bioinform.}
\bvolume{6}
\bpages{615--628}.
\bptok{imsref}%
\end{barticle}
\endbibitem

\bibitem{RasWil2006a}
\begin{bbook}[mr]
\bauthor{\bsnm{Rasmussen},~\bfnm{Carl~Edward}\binits{C.~E.}} \AND
  \bauthor{\bsnm{Williams},~\bfnm{Christopher K.~I.}\binits{C.~K.~I.}}
(\byear{2006}).
\btitle{Gaussian Processes for Machine Learning}.
\bpublisher{MIT Press}, \blocation{Cambridge, MA}.
\bid{mr={2514435}}
\bptok{imsref}%
\end{bbook}
\endbibitem

\bibitem{RegLijPru2003a}
\begin{barticle}[mr]
\bauthor{\bsnm{Regazzini},~\bfnm{Eugenio}\binits{E.}},
  \bauthor{\bsnm{Lijoi},~\bfnm{Antonio}\binits{A.}} \AND
  \bauthor{\bsnm{Pr{\"u}nster},~\bfnm{Igor}\binits{I.}}
(\byear{2003}).
\btitle{Distributional results for means of normalized random measures with
  independent increments}.
\bjournal{Ann. Statist.}
\bvolume{31}
\bpages{560--585}.
\bid{doi={10.1214/aos/1051027881}, issn={0090-5364}, mr={1983542}}
\bptok{imsref}%
\end{barticle}
\endbibitem

\bibitem{RicGre1997a}
\begin{barticle}[mr]
\bauthor{\bsnm{Richardson},~\bfnm{Sylvia}\binits{S.}} \AND
  \bauthor{\bsnm{Green},~\bfnm{Peter~J.}\binits{P.~J.}}
(\byear{1997}).
\btitle{On {B}ayesian analysis of mixtures with an unknown number of
  components}.
\bjournal{J. R. Stat. Soc. Ser. B Stat. Methodol.}
\bvolume{59}
\bpages{731--792}.
\bid{doi={10.1111/1467-9868.00095}, issn={0035-9246}, mr={1483213}}
\bptok{imsref}%
\end{barticle}
\endbibitem

\bibitem{Roe1994a}
\begin{barticle}[mr]
\bauthor{\bsnm{Roeder},~\bfnm{Kathryn}\binits{K.}}
(\byear{1994}).
\btitle{A graphical technique for determining the number of components in a
  mixture of normals}.
\bjournal{J. Amer. Statist. Assoc.}
\bvolume{89}
\bpages{487--495}.
\bid{issn={0162-1459}, mr={1294074}}
\bptok{imsref}%
\end{barticle}
\endbibitem

\bibitem{RoeWas97}
\begin{barticle}[mr]
\bauthor{\bsnm{Roeder},~\bfnm{Kathryn}\binits{K.}} \AND
  \bauthor{\bsnm{Wasserman},~\bfnm{Larry}\binits{L.}}
(\byear{1997}).
\btitle{Practical {B}ayesian density estimation using mixtures of normals}.
\bjournal{J.~Amer. Statist. Assoc.}
\bvolume{92}
\bpages{894--902}.
\bid{doi={10.2307/2965553}, issn={0162-1459}, mr={1482121}}
\bptok{imsref}%
\end{barticle}
\endbibitem

\bibitem{Ste00}
\begin{barticle}[mr]
\bauthor{\bsnm{Stephens},~\bfnm{Matthew}\binits{M.}}
(\byear{2000}).
\btitle{Bayesian analysis of mixture models with an unknown number of
  components---an alternative to reversible jump methods}.
\bjournal{Ann. Statist.}
\bvolume{28}
\bpages{40--74}.
\bid{doi={10.1214/aos/1016120364}, issn={0090-5364}, mr={1762903}}
\bptok{imsref}%
\end{barticle}
\endbibitem

\bibitem{titsmimak85}
\begin{bbook}[mr]
\bauthor{\bsnm{Titterington}, \bfnm{D.~M.}\binits{D.~M.}},
  \bauthor{\bsnm{Smith}, \bfnm{A.~F.~M.}\binits{A.~F.~M.}} \AND
  \bauthor{\bsnm{Makov}, \bfnm{U.~E.}\binits{U.~E.}}
(\byear{1985}).
\btitle{Statistical Analysis of Finite Mixture Distributions}.
\bpublisher{Wiley}, \blocation{Chichester}.
\bid{mr={0838090}}
\bptok{imsref}%
\end{bbook}
\endbibitem

\bibitem{TriFav2011a}
\begin{barticle}[mr]
\bauthor{\bsnm{Trippa},~\bfnm{Lorenzo}\binits{L.}} \AND
  \bauthor{\bsnm{Favaro},~\bfnm{Stefano}\binits{S.}}
(\byear{2012}).
\btitle{A class of normalized random measures with an exact predictive sampling
  scheme}.
\bjournal{Scand. J. Stat.}
\bvolume{39}
\bpages{444--460}.
\bid{doi={10.1111/j.1467-9469.2011.00749.x}, issn={0303-6898}, mr={2971631}}
\bptok{imsref}%
\end{barticle}
\endbibitem

\bibitem{Wal2007a}
\begin{barticle}[mr]
\bauthor{\bsnm{Walker},~\bfnm{Stephen~G.}\binits{S.~G.}}
(\byear{2007}).
\btitle{Sampling the {D}irichlet mixture model with slices}.
\bjournal{Comm. Statist. Simulation Comput.}
\bvolume{36}
\bpages{45--54}.
\bid{doi={10.1080/03610910601096262}, issn={0361-0918}, mr={2370888}}
\bptok{imsref}%
\end{barticle}
\endbibitem

\bibitem{WooBla2008a}
\begin{barticle}[auto:STB|2013/06/05|13:45:01]
\bauthor{\bsnm{Wood},~\bfnm{F.}\binits{F.}} \AND
  \bauthor{\bsnm{Black},~\bfnm{M.~J.}\binits{M.~J.}}
(\byear{2008}).
\btitle{A nonparametric Bayesian alternative to spike sorting}.
\bjournal{Journal of Neuroscience Methods}
\bvolume{173}
\bpages{1--12}.
\bptok{imsref}%
\end{barticle}
\endbibitem

\end{thebibliography}
\end{document}